\documentclass[12pt]{article}
\usepackage{amssymb}
\usepackage{amsmath,bm}
\usepackage{graphics,mathrsfs}
\usepackage{graphicx,epsfig}
\usepackage{subfigure}
\usepackage{setspace}
\usepackage{enumerate}
\usepackage{enumitem}
\usepackage{caption}
\usepackage{varioref}
\usepackage{color}
\usepackage{natbib}
\usepackage{epstopdf}
\setcitestyle{authoryear,open={(},close={)}}
\usepackage{amsthm}
\makeatletter \oddsidemargin  -.1in \evensidemargin -.1in
\begin{document}
	\newcommand{\bea}{\begin{eqnarray}}
	\newcommand{\eea}{\end{eqnarray}}
	\newcommand{\nn}{\nonumber}
	\newcommand{\bee}{\begin{eqnarray*}}
		\newcommand{\eee}{\end{eqnarray*}}
	\newcommand{\lb}{\label}
	\newcommand{\nii}{\noindent}
	\newcommand{\ii}{\indent}
	\newtheorem{thm}{Theorem}[section]
	\newtheorem{example}{Example}[section]
	\newtheorem{cor}{Corollary}[section]
	\newtheorem{definition}{Definition}[section]
	\newtheorem{lem}{Lemma}[section]
	\newtheorem{rem}{Remark}[section]
	\newtheorem{proposition}{Proposition}[section]
	\numberwithin{equation}{section}
	\renewcommand{\theequation}{\thesection.\arabic{equation}}
	\renewcommand\bibfont{\fontsize{10}{12}\selectfont}
	\setlength{\bibsep}{0.0pt}
\title{\bf Estimation of parameters of the Gumbel type-II distribution under AT-II PHCS with an application of Covid-$19$ data}
\author{ Subhankar {\bf Dutta}\thanks {Email address (corresponding author): subhankar.dta@gmail.com} ~and  Suchandan {\bf  Kayal}\thanks {Email address :
        kayals@nitrkl.ac.in,~suchandan.kayal@gmail.com}
    \\{\it \small Department of Mathematics, National Institute of
        Technology Rourkela, Rourkela-769008, India}}
\date{}
\maketitle

\begin{center}
 \textbf{Abstract}
\end{center}
In this paper, we investigate the classical and Bayesian estimation of unknown parameters of the Gumbel type-II distribution based on adaptive type-II progressive hybrid censored sample (AT-II PHCS). The maximum likelihood estimates (MLEs) and maximum product spacing estimates (MPSEs) are developed and computed numerically using Newton-Raphson method. Bayesian approaches are employed to estimate parameters under symmetric and asymmetric loss functions. Bayesian estimates are not in explicit forms. Thus, Bayesian estimates are obtained by using Markov chain Monte Carlo (MCMC) method along with the Metropolis-Hastings (MH) algorithm. Based on the normality property of MLEs the asymptotic confidence intervals are constructed. Also, bootstrap intervals and highest posterior density (HPD) credible intervals are constructed. Further a Monte Carlo simulation study is carried out. Finally, the data set based on the death rate due to Covid-19 in India is analyzed for illustration of the purpose. 

\section{Introduction}
In recent times, life testing experiments and reliability studies have achieved more acceptance. In such experiments, various situations appear where experimental units are eliminated before the time of failure. In these cases, the investigator may not have full information about the failure times of the whole experimental unit. Such kind of data collected from all these experiments are called censored data. It is worth mentioning that the censoring was introduced in practice to save time and reduce the number of failed units in a life-testing experiment.
Censoring can be done with respect to a prefixed time or prefixed number of failures or sometimes a combination of both prefixed time and number of failures. Depending upon these criteria, there are different types of censoring schemes. Let us consider a life testing experiment with $n$ number of experimental units and $X_{i:n}$ denotes the time of  $i$-th failure of the experimental units, where $i = 1,\ldots,n$. Let $T$ be a pre-specified time. If all the experiments fail before that pre-specified time $T$, it is known as a type-I censoring scheme. In the case of the type-II censoring scheme, the experiment will terminate after $m$-th number of failures observed, where $m \leq n$.
Both type-I and type-II censoring schemes have some disadvantages. In the case of type-I censoring, the number of failures may be zero, and in the case of type-II censoring, experimental time may be very large.
For further information about these schemes, we refer to \citet{lawless2011statistical}, \citet{sirvanci1984estimation} and \citet{balakrishnan1992estimation}. \\

\citet{epstein1954truncated} introduced a censoring scheme which is a mixture of type-I and type-II censoring schemes, known as the type-I hybrid censoring scheme. In this scheme, the experiment is terminated at time $T^{*}=~ min\{T, X_{m:n}\}$, where $X_{m:n}$ denotes the $m$-th failure and $T$ is pre-specified time. \citet{childs2003exact} proposed type-II hybrid censoring scheme where experiment will be terminated at time $T^{*}=max\{X_{m:n},T\}$.
For further studies on these hybrid censoring schemes, one may read \citet{balakrishnan2013hybrid}, \citet{kundu2007hybrid},  \citet{kundu2009estimating} and \citet{dey2014generalized}. In practical life testing experiments, there are many cases when the experimental units are removed from experiments before failure for other reasons. To overcome such scenarios, a type-II right censoring scheme is introduced, known as a progressive type-II censoring scheme. It has more flexibility in allowing the removal of units at time points other than the terminal point of the experiments.  For progressive type-II censoring scheme, let us consider that $n$ experimental units are placed on a life testing experiment. Let $X_{i:m:n}$ be the failure time of the $i$-th experimental unit, and $m$ be the number of failures that is fixed before the experiment starts. After first failure at $X_{1:m:n}$, $R_{1}$ number of units are removed randomly from $n-1$ surviving units. After second failure at $X_{2:m:n}$, $R_{2}$ number of units are removed randomly from $n-R_{1}-2$ surviving units. The test continues until $m$-th failure occurs and all remaining $R_{m} = n-\sum_{i=1}^{m}R_{i} -m$ surviving units are removed. In the progressive type-II censoring scheme, the values of $R_{i}$'s are pre-fixed. Many researchers have studied statistical inferences of different distributions under this censoring scheme. For example, see \citet{balakrishnan2003point}, \citet{rastogi2012estimating} and  \citet{maiti2019estimation}.  \\

In recent years, progressive type-II censoring scheme gained more popularity in life testing experiments and reliability studies. But there is a major drawback of this censoring scheme that it may take a large time to complete an experiment. To overcome this drawback,
\citet{kundu2006analysis} introduced progressive hybrid censoring scheme. Basically, there are two types of progressive hybrid censoring schemes. Let us consider  progressively censored ordered statistics  $X_{1:m:n},\ldots,X_{m:m:n}$ from a life testing experiments of $n$ units. Further, $(R_{1},\ldots,R_{m})$ are assumed to be the corresponding removals. If the experiment terminates at a time $T^{*}=~ min\{T,X_{m:m:n}\}$, where $T \in (0,\infty)$ is prefixed, then it is called type-I progressive hybrid censoring scheme (PHCS). In case of type-I PHCS,  at the time of first failure $X_{1:m:n}$, $R_{1}$ surviving units are removed randomly from $(n-1)$ units. At time of second failure $X_{2:m:n}$, $R_{2}$ number of surviving units are removed randomly from $(n-R_{1}-2)$ units. In similar way if $m$-th failure occurs before time $T$, i.e. $X_{m:m:n} < T$, then the observed failures are $X_{1:m:n},\ldots,X_{m:m:n}$ and all remaining units $R_{m}=n-m-\sum_{i=1}^{m-1}R_{i}$ are removed at termination point $X_{m:m:n}$. If $X_{m:m:n} > T,$ then the observed failures are $X_{1:m:n},\ldots,X_{j:m:n}$, where $X_{j:m:n} < T < X_{j+1:m:n}$ and all remaining units $R^{*}_{j}= n-j-\sum_{i=1}^{j}R_{i}$ at the time of termination $T$. In recent years, statistical inference under type-I PHC has been studied by several authors. One may refer to \citet{lin2012progressive}, \citet{tomer2015estimation} and \citet{kayal2017inference}. There is a drawback of type-I progressive hybrid censoring scheme, since the effective sample size may turn out to a small number. Thus, it has lower efficiency in computing statistical inferences for some problems.\\

To increase such efficiency, \citet{ng2009statistical} proposed adaptive type-II progressive hybrid censoring scheme (AT-II PHCS) where the effective sample size $m$ and time $T$ are pre-specified. This censoring scheme is similar to the type-I progressive hybrid censoring scheme. It is different when $T < X_{m:m:n}$. If $X_{j:m:n} < T < X_{j+1:m:n}$, where $j+1 < m$, all remaining units will be terminated by setting $R_{j+1}=\ldots=R_{m-1}=0$ and $R_{m}=n-m-\sum_{i=1}^{j}R_{j}$. Thus the experiment will be terminated after getting the expected sample size. In recent years many authors studied inferences of various lifetime models under AT-II PHCS. \citet{lin2009statistical} obtained classical estimates on the Weibull lifetime model based on AT-II PHCS. \citet{hemmati2013statistical} obtained maximum likelihood and approximated maximum likelihood estimates of model parameters of two parameters log-normal distribution based on AT-II PHCS. \citet{nassar2016estimation} discussed maximum likelihood estimates and Bayesian estimates of parameters of  Burr XII distribution model based on AT-II PHCS. \citet{panahi2020estimation} discussed classical and Bayesian estimation of parameters of inverted exponential Rayleigh distribution. \citet{panahi2021adaptive} discussed the maximum likelihood estimates and Bayesian estimates of parameters of Burr III distribution based on AT-II PHCS.  \\

\citet{gumbel2004statistics} introduced a two-parameter distribution, known as Gumbel type-II distribution, which is very useful to model meteorological phenomena such as floods, earthquakes, and natural disasters. Also it can used in life expectancy tables, hydrology and rainfall. The cumulative distribution function (CDF) of Gumbel type-II distribution is
 \begin{align}
 F(x)= e^{-\beta x^{-\alpha}}, ~~~x>0,~~\alpha,\beta >0, \label{eq 1.1}
 \end{align}
 where $\alpha$ is the shape parameter, $\beta$ is the scale parameter. The corresponding probability density function (PDF) is
 \begin{align}
 f(x)=\alpha\beta x^{-(\alpha+1)}e^{-\beta x^{-\alpha}},~~x>0,~~\alpha,\beta >0. \label{eq 1.2}
 \end{align}
 The hazard rate function of Gumbel type-II distribution is given by
 \begin{align}
 h(x)=\frac{\alpha \beta x^{-(\alpha +1)}}{e^{\beta x^{-\alpha}}-1},~~x>0,~~\alpha,\beta >0. \label{eq 1.3}
 \end{align}
\begin{figure}[h!]
 	\begin{center}
 		\subfigure[]{\label{c1}\includegraphics[height=2.5in,width=3.2in]{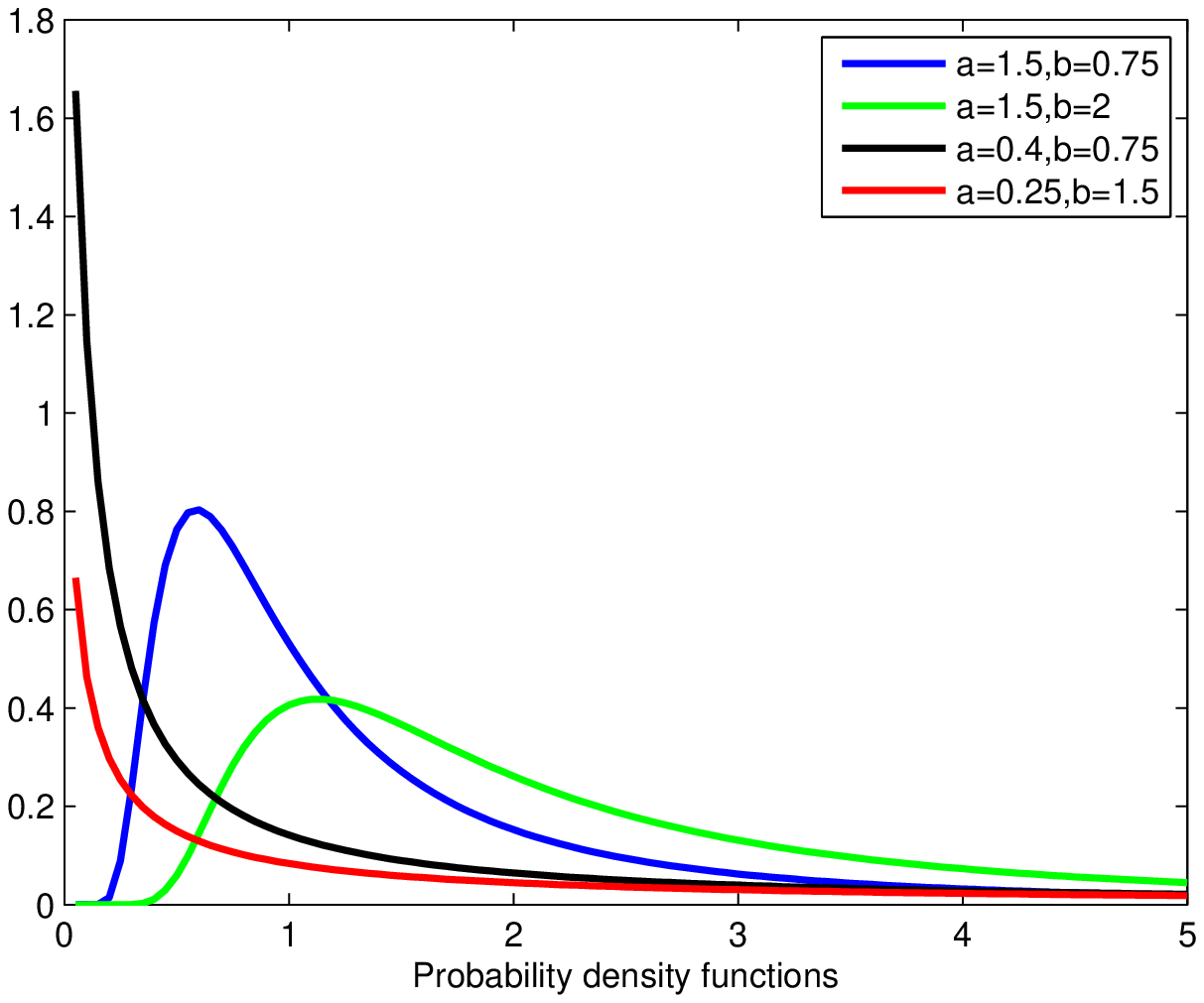}}
 		\subfigure[]{\label{c1}\includegraphics[height=2.5in,width=3.2in]{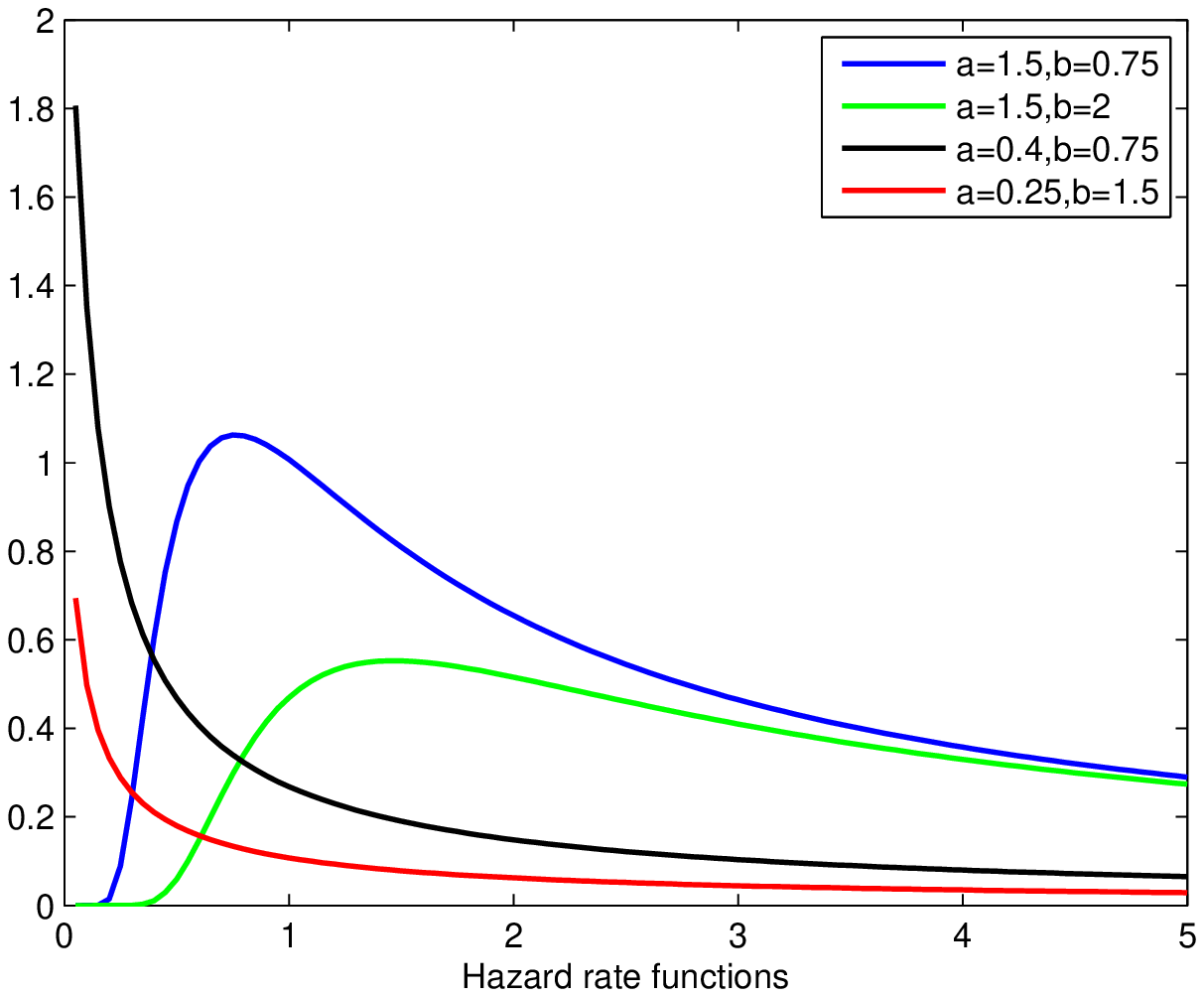}}\\
 		\caption{(a) PDF and (b) Hazard rate functions of Gumbel type-II distribution for different values of $\alpha$ and $\beta$.}
 	\end{center}
 \end{figure}
Figure $1$ represents graphs of the PDFs and hazard rate functions of the Gumbel type-II distribution based on different values of the parameters $\alpha$ and $\beta$. It is noticed that the shape of the hazard rate function of Gumbel type-II distribution is decreasing and upside-down bathtub (UTB). Due to these shapes of the hazard rate function, the Gumbel type-II distribution is very flexible to use in different research areas such as clinical, reliability and survival studies. Recently, many authors have studied statistical properties of the estimators of the model parameters of Gumbel type-II distribution. As example, \citet{abbas2013bayesian} discussed Bayesian estimation of the model parameters. \citet{reyad2015bayesian} studied E-Bayesian estimation of the unknown model shape parameter. \citet{sindhu2016study} obtained Bayes estimates and corresponding risk of the unknown model parameters based on left-censored data. \citet{abbas2020bayesian} proposed Bayesian estimation of the model parameters under type-II censored sample with some medical applications. \\

Due to importance of the Gumbel type-II distribution and the AT-II PHCS, in this paper, we have considered the problem of estimation of the parameters of Gumbel type-II distribution under AT-II PHCS. To the best of our information, this problem has not been studied yet. The rest of the article is organized as follows. In Section $2$, MLEs and MPSEs are computed by using the Newton-Raphson method. The Bayesian estimates are obtained under symmetric and asymmetric loss functions using the Markov chain Monte Carlo (MCMC) technique in Section $3$. In Section $4$, the asymptotic confidence intervals using the normality property of the MLEs and bootstrap confidence intervals are constructed. Also, MCMC samples are used to build HPD credible intervals. A Monte Carlo simulation study is carried out to compare the performance of different estimates in Section $5$. Section 6 deals with a real data set based on the death rates due to Covid-$19$ in India. Finally, conclusions of this paper have been drawn in Section $7$.

\section{Classical estimation}
In this section, classical estimation of unknown model parameters of Gumbel type-II distribution is obtained by using two methods: $(a)$ maximum likelihood estimation and $(b)$ maximum product spacing.

\subsection{Maximum likelihood estimation}
Let $x_{1:m:n} < \ldots <x_{j:m:n}<T<x_{j+1:m:n} < \ldots < x_{m:m:n}$ be an adaptive type-II progressive censored  ordered sample from $(\ref{eq 1.2})$  along with a censoring scheme $(R_{1},\ldots,R_{j},0,\ldots,0,R^{*}_{j})$, where $R^{*}_{j}=n-m-\sum_{i=1}^{j} R_{i}$  and $T$ is pre-specified. Then, the likelihood function can be written as
\begin{align}
L(\alpha,\beta | \underline{x}) \propto \prod_{i=1}^{m} f(x_{i:m:n}) \prod_{i=1}^{j} [1-F(x_{i:m:n})]^{R_{i}} [1-F(x_{m:m:n})]^{R^{*}_{j}}. \label{eq 2.1}
\end{align}
Whenever $x_{m:m:n} <T$, then $R^{*}_{j}=0$ and the likelihood function becomes
\begin{align}
L(\alpha,\beta | \underline{x}) \propto \prod_{i=1}^{m} f(x_{i:m:n}) [1-F(x_{i:m:n})]^{R_{i}}. \label{eq 2.2}
\end{align}
Replacing $F(x)$ and $f(x)$ from Equations $(\ref{eq 1.1})$ and $(\ref{eq 1.2})$ in Equation $(\ref{eq 2.1})$ we get,
\begin{align}
L \propto {\alpha}^m {\beta}^m \prod_{i=1}^{m}\bigg[{x_{i}}^{-(\alpha  +1)}  e^{-\beta{x_{i}^{-\alpha}}} \bigg] \prod_{i=1}^{j} \bigg[1-e^{-\beta{x_{i}^{-\alpha}}}\bigg]^{R_{i}} \bigg[1-e^{-\beta{x_{m}^{-\alpha}}}\bigg]^{R^{*}_{j}}, \label{eq 2.3}
\end{align}
where $x_{i}$ represents the $i$-th failure time $x_{i:m:n}$. The log-likelihood function is given as
\begin{align}
\nonumber \log L \propto &~~m \log {\alpha} + m \log {\beta} - (\alpha+1) \sum_{i=1}^{m}\log{x_{i}} - \beta \sum_{i=1}^{m}{x_{i}}^{-\alpha} + \sum_{i=1}^{j} R_{i} \log (1-e^{-\beta{x_{i}^{-\alpha}}})\\
& + R^{*}_{j} \log (1-e^{-\beta{x_{m}^{-\alpha}}}).  \label{eq 2.4}
\end{align}
After differentiating $(\ref{eq 2.4})$ partially with respect to $\alpha$ and $\beta$ respectively, and equating to zero we get
\begin{align}
\nonumber \frac{\partial l}{\partial \alpha} =&\frac{m}{\alpha} - \sum_{i=1}^{m} \log {x_{i}} + \beta \sum_{i=1}^{m}{x_{i}^{-\alpha}} \log {x_{i}} -\beta \sum_{i=1}^{j} \frac{R_{i} {x_{i}^{-\alpha}} e^{-\beta{x_{i}^{-\alpha}}}  \log {x_{i}} }{1-e^{-\beta{x_{i}^{-\alpha}}}} \\
 &- \beta \frac{R^{*}_{j} {x_{m}^{-\alpha}} e^{-\beta{x_{m}^{-\alpha}}}  \log {x_{m}} }{1-e^{-\beta{x_{m}^{-\alpha}}}} = 0,   \label{eq 2.5} \\
 \frac{\partial l}{\partial \beta} = &\frac{m}{\beta} -\sum_{i=1}^{m}{x_{i}^{-\alpha}} + \sum_{i=1}^{j} \frac{R_{i} {x_{i}^{-\alpha}} e^{-\beta{x_{i}^{-\alpha}}} }{1-e^{-\beta{x_{i}^{-\alpha}}}} +  \frac{R^{*}_{j} {x_{m}^{-\alpha}} e^{-\beta{x_{m}^{-\alpha}}} }{1-e^{-\beta{x_{m}^{-\alpha}}}} =0.  \label{eq 2.6}
\end{align}
The maximum likelihood estimates of unknown parameters $\alpha$ and $\beta$ can be obtained from Equations $(\ref{eq 2.5})$ and $(\ref{eq 2.6})$, which are not in closed form, thus need to be solved numerically. The maximum likelihood estimates of the unknown parameters are denoted by $\widehat{\alpha}$ and $\widehat{\beta}$. Further, when computing MLEs numerically, it is always of interest to study the existence and uniqueness of the MLEs of the parameters. In order to achieve this, one requires
to show two conditions proposed by \cite{makelainen1981existence}. These
are difficult to establish due to the complicated nature of the expressions
of the second order partial derivatives of the log-likelihood
function. To have a rough idea of that, the profile log-likelihood functions for the parameters $\alpha$ and $\beta$ are presented in Figure $2$. These figures represent that the MLEs may exist uniquely.
\begin{figure}[h!]
	\begin{center}
		\subfigure[]{\label{c1}\includegraphics[height=2in,width=3.2in]{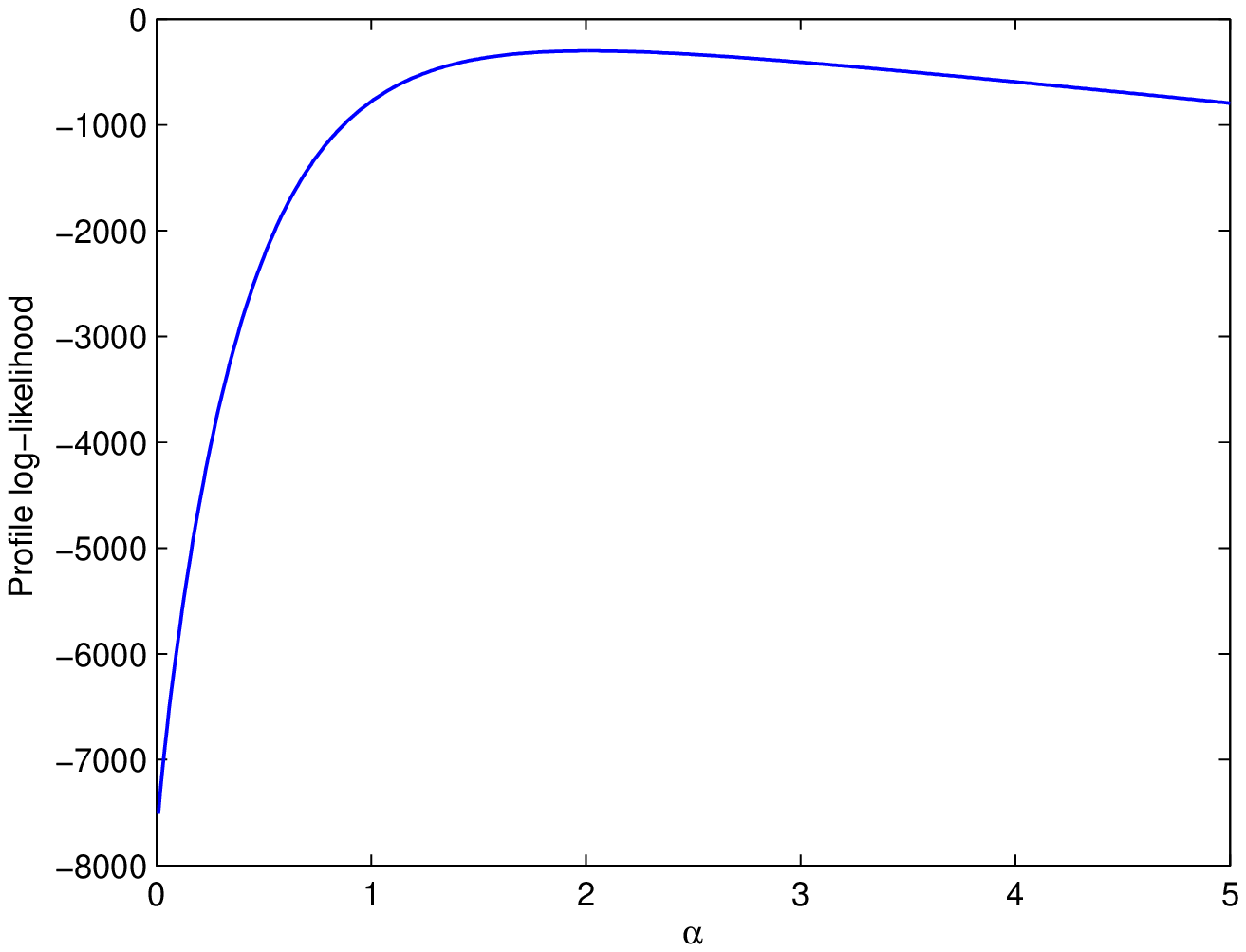}}
		\subfigure[]{\label{c1}\includegraphics[height=2in,width=3.2in]{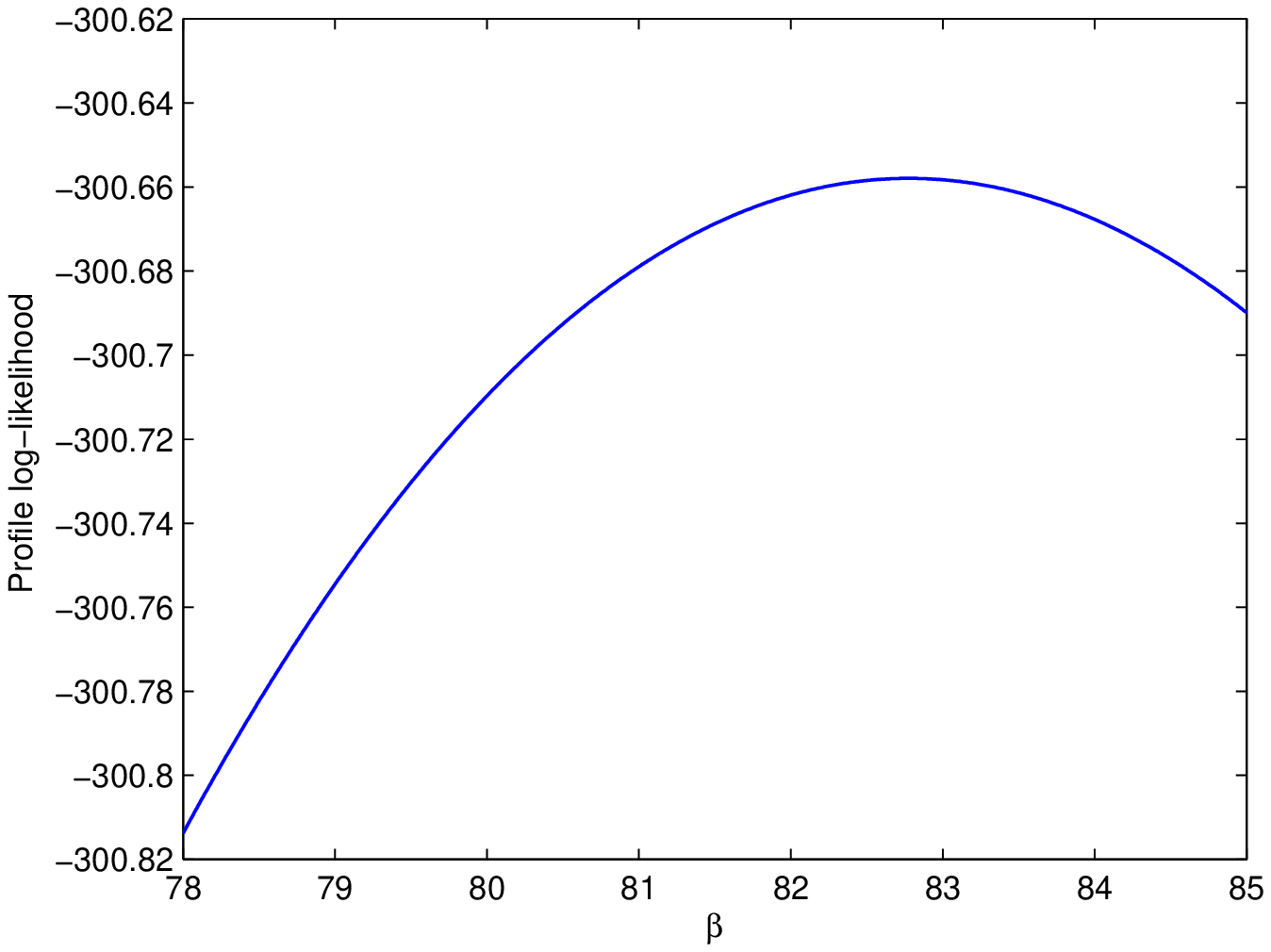}}
		\caption{Profile log-likelihood functions of $\alpha$ and $\beta$ for the real data set.}
	\end{center}
\end{figure}

\subsection{Maximum product spacing estimation }
\citet{cheng1983estimating} proposed maximum product spacing estimation (MPSE) method as an alternative to the MLE.
\citet{almetwally2019adaptive} discussed MPSE method under AT-II PHCS for generalized Rayleigh distribution. The product spacing under AT-II PHCS can be written as

\begin{align}
M = \prod_{i=1}^{m+1} D_{i} \prod_{i=1}^{j}\bigg[1-F(x_{i};\alpha,\beta)\bigg]^{R_{i}} \bigg[1-F(x_{m};\alpha,\beta)\bigg]^{R^{*}_{j}}, \label{eq 2.7}
\end{align}
where
 \begin{align}
 D_{i}= \begin{cases}
 F(x_{1}) ,& \text{ $i=1$ }\\
 F(x_{i})-F(x_{i-1}) ,& \text{ $i=2,3,\ldots,m$ }\\
 1-F(x_{m}) ,& \text{ $i= m+1$},
 \end{cases}
 \end{align}
 such that $\sum D_{i} =1$. Then, the product spacing function under AT-II PHCS based on Gumbel type-II model can be written as
 \begin{align}
 	M = e^{-\beta{x_{1}}^{-\alpha}} \bigg[1-e^{-\beta{x_{m}}^{-\alpha}}\bigg] \prod_{i=2}^{m} \bigg[ e^{-\beta{x_{i}}^{-\alpha}}- e^{-\beta{x_{i-1}}^{-\alpha}}\bigg] \bigg[1- e^{-\beta{x_{m}}^{-\alpha}}\bigg]^{R^{*}_{j}}.
 \end{align}
Now, the logarithm of space function can be written as
 \begin{align}
  \nonumber \log M =& {-\beta{x_{1}}^{-\alpha}} + \log (1-e^{-\beta{x_{m}}^{-\alpha}}) + \sum_{i=2}^{m} \log \big(e^{-\beta{x_{i}}^{-\alpha}}- e^{-\beta{x_{i-1}}^{-\alpha}}\big) \\
  & + \sum_{i-1}^{j} R_{i} \log (1-e^{-\beta{x_{i}}^{-\alpha}}) + R^{*}_{j} \log (1-e^{-\beta{x_{m}}^{-\alpha}}). \label{eq 2.10}
 \end{align}
 To obtain the normal equations from the above Equation $(\ref{eq 2.10})$, differentiate partially with respect to unknown parameters and equating to zero, we get \\
 \begin{align}
 \nonumber \frac{\partial \log M }{\partial \alpha }=&  -\frac{\beta {x_{m}}^{-\alpha} e^{-\beta {x_{m}}^{-\alpha}} }{(1-e^{-\beta {x_{m}}^{-\alpha}})} + \beta \sum_{i=2}^{m} \bigg[\frac{ {x_{i}}^{-\alpha}e^{-\beta {x_{i}}^{-\alpha}} \log {x_{i}}- {x_{i-1}}^{-\alpha}e^{-\beta {x_{i-1}}^{-\alpha}} \log {x_{i-1}} } {e^{-\beta {x_{i}}^{-\alpha}}-e^{-\beta {x_{i-1}}^{-\alpha}}} \bigg] \\
 & - \beta \sum_{i=1}^{j}  \frac{R_{i} {x_{i}}^{-\alpha}e^{-\beta {x_{i}}^{-\alpha}} \log {x_{i}} }{(1-e^{-\beta {x_{i}}^{-\alpha}})} -\frac{R^{*}_{j}\beta  {x_{m}}^{-\alpha}e^{-\beta {x_{m}}^{-\alpha}} \log {x_{m}} }{(1-e^{-\beta {x_{m}}^{-\alpha}})}+\beta {x_{1}}^{-\alpha} \log{x_{1}} =0.    \label{eq 2.11}
 \end{align}

 \begin{align}
 \nonumber \frac{\partial \log M }{\partial \beta }= & -{{x_{1}}^{-\alpha}}  + \sum_{i=2}^{m} \frac{{x_{i-1}}^{-\alpha}e^{-\beta {x_{i-1}}^{-\alpha}}-{x_{i}}^{-\alpha}e^{-\beta {x_{i}}^{-\alpha}}}{e^{-\beta {x_{i}}^{-\alpha}}-e^{-\beta {x_{i-1}}^{-\alpha}}} + \sum_{i=1}^{j} \frac{R_{i}{x_{i}}^{-\alpha}e^{-\beta {x_{i}}^{-\alpha}}}{(1-e^{-\beta {x_{i}}^{-\alpha}})} \\
 & + \frac{R^{*}_{j}{x_{m}}^{-\alpha}e^{-\beta {x_{m}}^{-\alpha}}}{(1-e^{-\beta {x_{m}}^{-\alpha}})}+ \frac{{x_{m}}^{-\alpha} e^{-\beta {x_{m}}^{-\alpha}}}{(1-e^{-\beta {x_{m}}^{-\alpha}})} =0.   \label{eq 2.12}
 \end{align}

 Above these two equations can not be solved analytically, so numerical method will be applied to obtain the MPSEs of the unknown parameters $\alpha$ and $\beta$, as $\widehat{\alpha}_{MPS}$ and $\widehat{\beta}_{MPS}$, respectively.

\section{Bayesian estimation}
In this section, Bayes estimates of unknown parameters of Gumbel type-II distribution are obtained with respect to symmetric and asymmetric loss functions under AT-II PHCS. To obtain the Bayes estimates of $\alpha$ and $\beta$, squared error loss function (SELF), which is symmetric and general entropy loss function (GELF) and LINEX loss function (LLF), which are asymmetric, are considered. Let $\tilde{\theta}$ be an estimator of $\theta$.  Then, SELF, LLF and GELF are respectively given by
\begin{align}
L_{SE}(\tilde{\theta},\theta) =~& (\tilde{\theta}-\theta)^2 \label{eq 3.1}    \\
L_{LI}(\tilde{\theta},\theta)=~& e^{p(\tilde{\theta}-\theta)}-p(\tilde{\theta}-\theta)-1,~ p\neq 0 \label{eq 3.2} \\
\nonumber \mbox{and}~~~~~~~~~~~~~~~~~~~~~~~~~~~~~~&\\
L_{GE}(\tilde{\theta},\theta)=~& \bigg(\frac{\tilde{\theta}}{\theta}\bigg)^q -q\log \bigg(\frac{\tilde{\theta}}{\theta}\bigg)-1,~ q \neq 0. \label{eq 3.3}
\end{align}
Under the loss functions given by (\ref{eq 3.1}), (\ref{eq 3.2}) and (\ref{eq 3.3}), the Bayes estimates of $\theta$ can be respectively written as
\begin{align}
\widehat{\theta}_{SE}=& E_{\theta}(\theta |\underline{x}), \label{eq 3.4} \\
\widehat{\theta}_{LI}=& -p^{-1} \log[E_{\theta}(e^{-p\theta}|\underline{x})],~~ \label{eq 3.5}  \\
\nonumber \mbox{and}~~~~~~~~~~~~~~~~~~~~~~~~~~~&\\
\widehat{\theta}_{GE}=& [E_{\theta}({\theta}^{-q}|\underline{x})],~  q \neq 0.    \label{eq 3.6}
\end{align}
In Bayesian estimation, choosing priors for the unknown model parameters is an important as well as challenging problem. There is no clear methodology to choose best priors for Bayesian estimation problem. Here, we assume the prior distributions for the parameters as $\alpha \sim Gamma(a,b)$ and $\beta \sim Gamma(c,d)$, where $Gamma(a,b)$ denotes gamma distribution with shape parameter $a$ and scale parameter $b$. The joint prior density of $\alpha$ and $\beta$ can be written as
\begin{align}
\pi^{*}(\alpha,\beta)= {\alpha}^{a-1}{\beta}^{c-1} e^{-(b\alpha +d\beta)},~~\alpha,\beta>0,~a,b,c,d>0. \label{eq 3.7}
\end{align}
Based on the likelihood function (\ref{eq 2.3}) and the joint prior density function (\ref{eq 3.7}), the posterior density function of $\alpha$ and $\beta$ can be written as
\begin{eqnarray}
\nonumber \pi(\alpha,\beta |\underline{x})&=&k^{-1} {\alpha}^{m+a-1} e^{-\alpha(b+\sum_{i=1}^{m}\log x_{i})}{\beta}^{m+c-1} e^{-\beta(d+\sum_{i=1}^{m}{x_{i}}^{-\alpha})} \big(1-e^{-\beta {x_{m}}^{-\alpha}}\big)^{R^{*}_{j}} \\
&~&\times \prod_{i=1}^{j}\big(1-e^{-\beta {x_{i}}^{-\alpha}}\big)^{R_{i}}, \label{eq 3.8}
\end{eqnarray}
where
\begin{eqnarray}
\nonumber k&=&~ \int_{0}^{\infty}\int_{0}^{\infty}k^{-1} {\alpha}^{m+a-1} e^{-\alpha(b+\sum_{i=1}^{m}\log x_{i})}{\beta}^{m+c-1} e^{-\beta(d+\sum_{i=1}^{m}{x_{i}}^{-\alpha})} \big(1-e^{-\beta {x_{m}}^{-\alpha}}\big)^{R^{*}_{j}} \\
&~&\times \prod_{i=1}^{j}\big(1-e^{-\beta {x_{i}}^{-\alpha}}\big)^{R_{i}}~ d\alpha d\beta
\end{eqnarray}
and all the hyper-parameters $a,b,c,d$ are non-negative and known. Now, consider a function of parameters $\alpha$ and $\beta$, say $\psi(\alpha,\beta)$. Then, from (\ref{eq 3.4}), (\ref{eq 3.5}) and (\ref{eq 3.6}), the Bayes estimates of $\psi(\alpha,\beta)$ with respect to SELF, LLF and GELF are given by
\begin{eqnarray}
\widehat{\psi}_{SE}&=&~ \int_{0}^{\infty}\int_{0}^{\infty} \psi(\alpha,\beta) \pi(\alpha,\beta |\underline{x})~ d\alpha d\beta, \label{eq 3.10}\\
\widehat{\psi}_{LI}&=&~-\bigg(\frac{1}{p}\bigg)\log \bigg[ \int_{0}^{\infty}\int_{0}^{\infty}e^{-p\psi(\alpha,\beta)} \pi(\alpha,\beta |\underline{x})~ d\alpha d\beta\bigg], \label{eq 3.11} \\
\nonumber \mbox{and}~~~~~~~~~~~~~~~~~~~~~&\\
\widehat{\psi}_{GE}&=&~ \bigg[\int_{0}^{\infty}\int_{0}^{\infty} (\psi(\alpha,\beta))^{-q} \pi(\alpha,\beta |\underline{x})~ d\alpha d\beta\bigg]^{-\frac{1}{q}}.   \label{eq 3.12}
\end{eqnarray}
respectively. To derive Bayes estimates of $\alpha$ and $\beta$ in respect of loss functions given by (\ref{eq 3.1}), (\ref{eq 3.2}) and (\ref{eq 3.3}), $\psi(\alpha,\beta)$ is replaced by $\alpha$ and $\beta$ respectively in Equations (\ref{eq 3.10}), (\ref{eq 3.11}) and (\ref{eq 3.12}). Then, the Bayes estimates of $\alpha$ are given by
\begin{eqnarray}
\nonumber \widehat{\alpha}_{SE}&=&~ {k_{1}}^{-1} \int_{0}^{\infty}\int_{0}^{\infty}{\alpha}^{m+a} e^{-\alpha(b+\sum_{i=1}^{m}\log x_{i})}{\beta}^{m+c-1} e^{-\beta(d+\sum_{i=1}^{m}{x_{i}}^{-\alpha})} \\
&~&\times  \big(1-e^{-\beta {x_{m}}^{-\alpha}}\big)^{R^{*}_{j}} \prod_{i=1}^{j}\big(1-e^{-\beta {x_{i}}^{-\alpha}}\big)^{R_{i}}~ d\alpha d\lambda, \label{eq 3.13} \\
\nonumber \widehat{\alpha}_{LI}&=&~ -\bigg(\frac{1}{p}\bigg) \log \bigg[{k_{1}}^{-1}\int_{0}^{\infty}\int_{0}^{\infty}{\alpha}^{m+a-1} e^{-\alpha(b+p+\sum_{i=1}^{m}\log x_{i})}{\beta}^{m+c-1} e^{-\beta(d+\sum_{i=1}^{m}{x_{i}}^{-\alpha})}  \\
&~&\times \big(1-e^{-\beta {x_{m}}^{-\alpha}}\big)^{R^{*}_{j}} \prod_{i=1}^{j}\big(1-e^{-\beta {x_{i}}^{-\alpha}}\big)^{R_{i}}~ d\alpha d\lambda\bigg], \label{eq 3.14} \\
\nonumber \mbox{and}~~~~~~~~~~&\\
\nonumber \widehat{\alpha}_{GE}&=&~ \bigg[{k_{1}}^{-1}\int_{0}^{\infty}\int_{0}^{\infty} {\alpha}^{m+a-q-1} e^{-\alpha(b+\sum_{i=1}^{m}\log x_{i})}{\beta}^{m+c-1} e^{-\beta(d+\sum_{i=1}^{m}{x_{i}}^{-\alpha})}  \\
&~&\times \big(1-e^{-\beta {x_{m}}^{-\alpha}}\big)^{R^{*}_{j}} \prod_{i=1}^{j}\big(1-e^{-\beta {x_{i}}^{-\alpha}}\big)^{R_{i}}~ d\alpha d\lambda \bigg]^{-\frac{1}{q}}.    \label{eq 3.15}
\end{eqnarray}
In a similar way, by replacing $\psi(\alpha,\beta)$ as $\beta$ in Equations (\ref{eq 3.10}), (\ref{eq 3.11}) and (\ref{eq 3.12}) Bayes estimates $\widehat{\beta}_{SE}$, $\widehat{\beta}_{LI}$ and $\widehat{\beta}_{GE}$ can be obtained. There are ratios of two integrals given in Equations $(\ref{eq 3.13})$, $(\ref{eq 3.14})$ and $(\ref{eq 3.15})$ which can not be obtained in a closed form. To overcome such situations, MCMC technique will be used to obtain desired Bayes estimates.

\subsection{MCMC method}
In this section, the Bayes estimates of the unknown parameters $\alpha$ and $\beta$ are computed. To generate samples from $(\ref{eq 3.8})$, MCMC method has been used. From posterior probability density function given by $(3.8)$, the conditional posterior densities of the parameters can be written as
\begin{eqnarray}
\nonumber\pi_{1}(\alpha|\beta,\underline{x})&\propto~ &{\alpha}^{m+a-1} e^{-\alpha(b+\sum_{i=1}^{m}\log x_{i})} e^{-\beta(d+\sum_{i=1}^{m}{x_{i}}^{-\alpha})} \prod_{i=1}^{D}(1-e^{-\beta{x_{i}}^{-\alpha}})^{R_{i}}\\
 &~& \times (1-e^{-\beta{x_{m}}^{-\alpha}})^{R^{*}_{j}} \label{eq 3.16}
\end{eqnarray}
and
\begin{eqnarray}
\pi_{2}(\beta|\alpha,\underline{x})&\propto~& {\beta}^{m+c-1}  e^{-\beta(d+\sum_{i=1}^{m}{x_{i}}^{-\alpha})} \prod_{i=1}^{D}(1-e^{-\beta{x_{i}}^{-\alpha}})^{R_{i}} (1-e^{-\beta{x_{m}}^{-\alpha}})^{R^{*}_{j}}. \label{eq 3.17}
\end{eqnarray}
Note that the density functions $\pi_{1}(\alpha|\beta,\underline{x})$ and $\pi_{2}(\beta|\alpha,\underline{x})$ can not be brought into some well known classes of distributions. Thus, the MCMC samples for $\alpha$ and $\beta$ can not be generated directly. In this context, the Metropolis-Hastings algorithm (see \citet{chen2012monte}) is used to generate samples from $(\ref{eq 3.16})$ and $(\ref{eq 3.17})$. This method has been discussed  as follows: \\\\
\textbf{Step 1}~ Set $i=1$ and choose initial guesses as ${\alpha}^{(1)}= \widehat{\alpha}$ and $\beta^{(1)}= \widehat{\beta}$. \\
\textbf{Step 2}~ Generate new samples  and ${\beta}^{(i)}$ with proposal distribution ${\alpha}^{(i)} \sim N ({\alpha}^{(i-1)},var(\widehat{\alpha}))$ and ${\beta}^{(i)} \sim N ({\beta}^{(i-1)},var(\widehat{\beta}))$.\\
\textbf{Step 3}~ Compute $h= min\{1,\frac{\pi({\alpha}^{(i)},{\beta}^{(i)}|\underline{x})}{\pi({\alpha}^{(i-1)},{\beta}^{(i-1)}|\underline{x}) }\}$. \\
\textbf{Step 4}~ Generate a sample $u$ from $U(0,1)$. \\
\textbf{Step 5}~ Set $(\alpha,\beta)=({\alpha}^{(i)},{\beta}^{(i)})$, if $u \leq h$; otherwise $(\alpha,\beta)=({\alpha}^{(i-1)},{\beta}^{(i-1)})$.\\
\textbf{Step 6}~ Set $i=i+1$.\\
\textbf{Step 7}~ Repeat Steps (2-6) $N$ number of times to get ${\alpha}^{(1)},\ldots,{\alpha}^{(N)}$ and ${\beta}^{(1)},\ldots,{\beta}^{(N)}$. \\\\
Then, the Bayes estimates of the parameters under SELF are given by
\begin{align}
\widehat{\alpha}_{SE}= \frac{1}{N} \sum_{i=1}^{N} {\alpha}^{(i)}~~ \mbox{and}~~ \widehat{\beta}_{SE}= \frac{1}{N} \sum_{i=1}^{N} {\beta}^{(i)}. \label{eq 3.18}
\end{align}
The Bayes estimates of the parameters under LLF are obtained as
\begin{align}
\widehat{\alpha}_{LI}= -\bigg(\frac{1}{p}\bigg)\log\bigg(\frac{1}{N} \sum_{i=1}^{N} e^{-p{\alpha}^{(i)}}\bigg) ~~\mbox{and}~~ \widehat{\beta}_{LI}= -\bigg(\frac{1}{p}\bigg)\log\bigg(\frac{1}{N} \sum_{i=1}^{N} e^{-p{\beta}^{(i)}}\bigg),~~p\ne0.     \label{eq 3.19}
\end{align}
Further, 
the Bayes estimates of the parameters under GELF are proposed as
\begin{align}
\widehat{\alpha}_{GE}= \bigg[\frac{1}{N} \sum_{i=1}^{N} ({\alpha}^{(i)})^{-q}\bigg]^{-q}~~ \mbox{and}~~ \widehat{\beta}_{GE}= \bigg[\frac{1}{N} \sum_{i=1}^{N} ({\beta}^{(i)})^{-q}\bigg]^{-q},~~q\ne0. \label{eq 3.20}
\end{align}

\section{Confidence intervals}
In this section, three types of confidence intervals for $\alpha$ and $\beta$, namely asymptotic confidence intervals, bootstrap confidence intervals and highest posterior density intervals are constructed.
\subsection{Asymptotic confidence interval}
The $100(1-\gamma) \%$ asymptotic confidence intervals for $\alpha$ and $\beta$ can be constructed by using asymptotic normality property of the MLEs $\widehat{\alpha}$ and $\widehat{\beta}$. In doing so, variance of $\widehat{\alpha}$ and $\widehat{\beta}$ are required. These can be obtained from main diagonal elements of the inverse of the observed Fisher information matrix $\widehat{I}^{-1}(\widehat{\alpha},\widehat{\beta})$, where 
\begin{align}\label{3.21}
\widehat{I}(\widehat{\alpha},\widehat{\beta})= {\begin{bmatrix}
	-l_{20} & -l_{11}\\
	-l_{11} & -l_{02}\\
	\end{bmatrix}}_{({\alpha},{\beta})=(\widehat{\alpha},\widehat{\beta})}
\end{align}
and $l_{ij}= \frac{\partial^2 logL}{\partial \theta_{i}\partial \theta_{j}}$, $\Theta=(\theta_{1},\theta_{2})= (\alpha,\beta)$. Therefore, the $100(1-\gamma) \%$ approximate confidence intervals for $\alpha$ and $\beta$ are respectively given by
\begin{align}
\nonumber \bigg(\widehat{\alpha}~ \underline{+}~z_{\frac{\gamma}{2}}\sqrt{Var(\widehat{\alpha})}\bigg) ~~\mbox{and}~~ \bigg(\widehat{\beta}~ \underline{+}~z_{\frac{\gamma}{2}}\sqrt{Var(\widehat{\beta})}\bigg),
\end{align}
where $z_{\frac{\gamma}{2}}$ is the upper $\frac{\gamma}{2}$-th percentile point of a standard normal distribution.

\subsection{Bootstrap confidence interval}
The approximate confidence intervals are acceptable when the effective sample size $m$ is large. When $m$ is small, efficiency of construction of the asymptotic confidence intervals may not work properly. Bootstrap re-sampling may produce approximate confidence intervals with more accuracy. Here, two types of parametric bootstrap confidence intervals are constructed.\\\\
\textbf{4.2.1 Percentile bootstrap (Boot-p) confidence interval}\\\\
\textbf{Step 1} Compute $\widehat{\alpha}$ and $\widehat{\beta}$.\\
\textbf{Step 2} Use these $\widehat{\alpha}$, $\widehat{\beta}$ and same $T$, $R_{i}$'s and $m$ to generate a bootstrap re-sample. \\
\textbf{Step 3} Obtain bootstrap estimates $\widehat{\alpha}^{B}$  and $\widehat{\beta}^{B}$ from these bootstrap sample. \\
\textbf{Step 4} Repeat Steps 2-3 up to $N$ times to get  $\widehat{\alpha}^{B[1]},\ldots,\widehat{\alpha}^{B[N]}$ and $\widehat{\beta}^{B[1]},\ldots,\widehat{\beta}^{B[N]}$.\\
\textbf{Step 5} Rearrange these bootstrap estimates in ascending order as $\widehat{\alpha}^{B(1)},\ldots,\widehat{\alpha}^{B(N)}$ and $\widehat{\beta}^{B(1)},\ldots,\widehat{\beta}^{B(N)}$.\\\\
The $100(1-\gamma)\%$ percentile bootstrap confidence intervals for $\alpha$ and $\beta$ are respectively given by
\begin{align}
\nonumber \bigg(\widehat{\alpha}^{B(N\gamma/2)},\widehat{\alpha}^{B(N(1-\gamma/2)}\bigg)~~\mbox{and}~~\bigg(\widehat{\beta}^{B(N\gamma/2)},\widehat{\beta}^{B(N(1-\gamma/2)}\bigg).
\end{align}
\textbf{4.2.2 Bootstrap-t (Boot-t) confidence interval}\\\\
Steps 1-2 are similar to above discussed Boot-p method. Thus, we only state Steps $3$ and $4$ below: \\
\textbf{Step 3} Set t-statistics as $T_{1}= \frac{\widehat{\alpha}^{B}-\alpha}{\sqrt{Var(\widehat{\alpha}^{B})}}$ and $T_{2}= \frac{\widehat{\beta}^{B}-\beta}{\sqrt{Var(\widehat{\beta}^{B})}}$ and then compute. \\
\textbf{Step 4} Repeat Steps 2-3 up to $N$ times to get ${T_{1}}^{[1]},\ldots,{T_{1}}^{[N]}$ and ${T_{2}}^{[1]},\ldots,{T_{2}}^{[N]}$ and then rearranging these to get ${T_{1}}^{(1)},\ldots,{T_{1}}^{(N)}$ and ${T_{2}}^{(1)},\ldots,{T_{2}}^{(N)}$. \\\\
Then, $100(1-\gamma)\%$ bootstrap-t confidence intervals are obtained as
\begin{align}
\nonumber \bigg({T_{1}}^{(N\gamma/2)},{T_{1}}^{(N(1-\gamma/2))}\bigg)~~\mbox{and}~~\bigg({T_{2}}^{(N\gamma/2)},
{T_{2}}^{(N(1-\gamma/2))}\bigg), 
\end{align}
respectively. 

\subsection{HPD credible intervals}
In this section, using MCMC samples ${\alpha}^{(1)},\ldots,{\alpha}^{(N)}$ and ${\beta}^{(1)},\ldots,{\beta}^{(N)}$ and the method given by \citet{chen2012monte}, HPD credible intervals for model parameters are obtained. After ordering the MCMC samples in increasing way, these samples can be written as ${\alpha}_{(1)},\ldots,{\alpha}_{(N)}$ and  ${\beta}_{(1)},\ldots,{\beta}_{(N)}$. 
The $100(1-\gamma/2)\%$ credible intervals are obtained as
   \begin{align}
   \nonumber ({\alpha}_{(k)}, {\alpha}_{(k+(1-\gamma)N)})~~\mbox{and}~~({\beta}_{(k)}, {\beta}_{(k+(1-\gamma)N)})
   \end{align}
   where $k=1,\ldots,[\gamma N]$, where $[\cdot]$ represent the greatest integer function and the corresponding interval lengths are $l_{1k}= {\alpha}_{(k+(1-\gamma)N)}- {\alpha}_{(k)}$ and $l_{2k}= {\beta}_{(k+(1-\gamma)N)}- {\beta}_{(k)}$ . Then, put away these intervals for which  $l_{1k}$ and $l_{2k}$ become the smallest to obtain the HPD credible intervals.\\

\section{Simulation study}
In this section, a Monte Carlo simulation study is carried out to compare the performance of different estimates of parameters for Gumbel type-II distribution based on adaptive type-II progressive hybrid censoring schemes. The performance of all estimates has been compared in terms of their absolute bias (AB) and mean squared errors (MSEs). For this purpose, $10000$ AT-II progressive hybrid censored samples are generated by using different values of $n, m, T$ along with three following censoring schemes:\\[0.2 cm]
Scheme 1 : $R_{1}=\ldots=R_{m-1}=0,R_{m}= n-m$. \\
Scheme 2 : $R_{1}=n-2m+1,R_{2}=\ldots=R_{m}=1$.\\
Scheme 3 : $R_{1}=n-m-5,R_{2}=\ldots=R_{m-5}=0,R_{m-4}=\ldots=R_{m}=1$.\\[0.2 cm]
The average bias and MSEs of the MLEs, MPSEs and Bayesian estimates are computed for $\alpha=1.5$ and $\beta=0.75$. These are presented in Table {\ref{Table 1}}, Table {\ref{Table 2}} , Table \ref{Table 4} and Table \ref{Table 5}. The Bayesian estimates are computed by using MCMC method along with 5000 MCMC samples. The hyper parameters in gamma prior are considered as $a=3,b=2,c=3,$ and $d=4$. Further, the 95$\%$ confidence interval lengths and coverage probabilities for asymptotic confidence intervals, bootstrap intervals and HPD credible intervals are computed and tabulated in Table $\ref{Table 3}$ and Table $\ref{Table 6}$.
From the tables, the following conclusions are made :
\begin{itemize}
 \item [(i)] For fixed values of $n$ and $T$, when value of $m$ increases, then the values of absolute bias and MSEs of MLEs, MPSEs and the Bayes estimates decrease.
\item [(ii)] For different values of $n$ and $T$, when $m=10$, then Scheme $2$ performes better than other two schemes; but when $m=15$, then Scheme $3$ performes better than other two schemes.
\item [(iii)] In most of the cases, for fixed values of $n$ and $m$, when $T$ deceases the values of MSEs of the estimates increase.
\item [(iv)] For fixed values of $n$ and $T$, when value of $m$ increases the values of the average lengths increase.
\item [(v)] It have been noticed that the Bayes estimates perform better than the classical estimates in terms of the absolute bias and MSEs. In classical estimates, MPSEs perform better than MLEs in terms of the absolute bias and MSEs.
\item [(vi)] Bayes estimates with respect to the LLF when $p=0.25$ provide superior performance than other Bayes estimates.
\item [(vii)] The performance of the HPD credible intervals is better than other confidence intervals in the sense of the average lengths and coverage probabilities.
\end{itemize}

\begin{table}[!htbp]
	\renewcommand\thetable{1}
	\scriptsize \caption{\label{Table 1}Absolute bias and MSE of
		estimates of $\alpha$ when $T=1.5$.}
	\scalebox{0.95}{
	\begin{tabular}{cccccccccccccccccc}
		\hline\\
		(n,m)& scheme  &$\widehat{\alpha}$& &$\widehat{\alpha}_{MPS}$  && &&  $\widehat{\alpha}_{LI}$& & &  &$\widehat{\alpha}_{GE}$& &  &$\widehat{\alpha}_{SE}$ \\
		\hline	
		& & & && & &$p=-0.25$ & &$p=0.25$& &  $q=-0.25$ && $q=0.25$  && \\
		\hline\\
		(30,10)& I&0.5098 &&0.3317& & &0.1786 & &0.1772& &  0.1778 && 0.1780  && 0.1778\\
		& &0.3336 && 0.1986& & &0.0501 & &0.0489& &  0.0492  && 0.0490  && 0.0495\\
		& II& 0.3658  &&0.2987 & & &0.1803 & &0.1789& &  0.1795 && 0.1796  && 0.1795 \\
		& &0.2691 &&0.1588 & & &0.0508 & &0.0497& & 0.0499 && 0.0498 && 0.0503\\
		& III&0.3563 & &0.2919& & &0.1765 & &0.1754& & 0.1761  &&  0.1765  && 0.1759\\
		& &0.2490 & &0.1479& & &0.0492 & &0.0482& &  0.0485 && 0.0484 &&0.0486 \\
		\hline\\
		(30,15)&I &0.3191 & &0.2568& & &0.1701 & &0.1687& &  0.1692 && 0.1693  && 0.1694 \\
		& & 0.3577 & &0.1085& & &0.0458 & &0.0446& &  0.0448 && 0.0446 &&  0.0452\\
		& II&0.3053 &&0.2747 & & &0.1732 & &0.1723& &  0.1731 && 0.1735  && 0.1727\\
		& &0.2005 &&0.1168 & & &0.0473 & &0.0465& &  0.0468 && 0.0469  &&0.0469 \\
		&III&0.2913  &&0.2779& & &0.1756 & &0.1751& &  0.1761 && 0.1767  &&0.1753 \\
		& &0.1617  &&0.1174 & & &0.0484 & &0.0478& & 0.0482 && 0.0485  && 0.0481\\
		\hline\\
		(40,10)&I&0.4630 & &0.3317& & &0.1688 & &0.1675& &  0.1679 && 0.1680  && 0.1681\\
		& &0.9015 & &0.1990& & &0.0451 & &0.0439& &  0.0441 && 0.0439  && 0.0445\\
		& II&0.3432 && 0.2854& & &0.1710 & &0.1696& &  0.1700 &&  0.1699 && 0.1703\\
		& &0.2281 &&0.1437 & & &0.0458  & &0.0447& &  0.0448 && 0.0447  && 0.0453\\
		& III&0.3344 & &0.2797& & &0.1708 & &0.1696& &  0.1702 && 0.1703  &&0.1702 \\
		& & 0.2124 & &0.1367& & &0.0455& &0.0455 & & 0.0447 && 0.0446  && 0.0450 \\
		\hline\\
		(40,15)& I&0.3484 & &0.2588& & &0.1641 & & 0.1628& &  0.1631 && 0.1630  && 0.1634\\
		& &0.5966 & &0.1110& & &0.0427 & &0.0416& &  0.0417 && 0.0415  && 0.0422 \\
		& II&0.2876 & &0.2627& & &0.1647 & &0.1637& &  0.1643 && 0.1645  && 0.1642\\
		& &0.1702 & &0.1066& & &0.0429 & &0.0421& &  0.0423 && 0.0423  && 0.0425 \\
		& III&0.2745 & &0.2666& & &0.1668 & &0.1661& &  0.1669 && 0.1667  && 0.1664\\
		& &0.1353 & &0.1081& & &0.0437 & &0.0431& & 0.0434 && 0.0435  && 0.0434\\
		\hline
   \end{tabular}}
\end{table}

\begin{table}[!htbp]
	\renewcommand\thetable{2}
	\scriptsize \caption{\label{Table 2}Absolute bias and MSE of
		estimates of $\beta$ when $T=1.5.$}
	\scalebox{0.95}{
	\begin{tabular}{cccccccccccccccccc}
		\hline\\
		(n,m)& scheme  &$\widehat{\beta}$& &$\widehat{\beta}_{MPS}$  && &&  $\widehat{\beta}_{LI}$& & &  &$\widehat{\beta}_{GE}$& &  &$\widehat{\beta}_{SE}$ \\
		\hline	
		& & & && & &$p=-0.25$ & &$p=0.25$& &  $q=-0.25$ && $q=0.25$  && \\
		\hline\\
		(30,10)& I&0.9689 &&0.1753& & &0.0827 & &0.0826& &   0.0828 && 0.0831  && 0.0826\\
		& &0.5365 & &0.0479& & &0.0108 & &0.0107& &  0.0107 && 0.0108  && 0.0107\\
		& II&0.2237  &&0.1899 & & &0.0859 & &0.0858& &  0.0862&& 0.0865  && 0.0859 \\
		& &0.1384 &&0.0575 & & &0.0115 & &0.0114& &  0.0115 && 0.0116 && 0.0115\\
		& III&0.2251 & &0.1919& & &0.0852 & &0.0850& &  0.0854 && 0.0857 && 0.0851\\
		& &0.1251 & &0.0582& & &0.0114 & &0.0113& &  0.0114 && 0.0115  && 0.0114\\
		\hline\\
		(30,15)&I &0.2694 & &0.1414& & &0.0818 & &0.0816& &  0.0819 && 0.0821  && 0.0817 \\
		& &0.2713 & &0.0319& & &0.0106 & &0.0105& &  0.0105 && 0.0106  && 0.0105\\
		& II&0.1815 &&0.1586 & & &0.0835 & &0.0834& &  0.0836 && 0.0838  &&0.0835 \\
		& &0.0855 &&0.0404 & & &0.0110 & &0.0109& & 0.0109 && 0.0110  && 0.0109\\
		&III&0.1879  &&0.1690& & &0.0848 & &0.0846& &  0.0849 && 0.0850  && 0.0847 \\
		& &0.0669  &&0.0459 & & &0.0113 & &0.0112& & 0.0113 && 0.0114  && 0.0113 \\
		\hline\\
		(40,10)&I&1.2630 & &0.1804& & &0.0826 & & 0.0825& &  0.0828 && 0.0831  && 0.0826 \\
		& &0.7557 & &0.0505& & &0.0107 & &0.0106& &  0.0106 && 0.0107  && 0.0106\\
		& II&0.2193 & &0.1923& & &0.0859 & &0.0858& &  0.0861 && 0.0863  &&0.0858\\
		& &0.1234 & &0.0589& & &0.0114 & &0.0113& &  0.0115 && 0.0116  && 0.0114\\
		& III&0.2218 & &0.1965& & &0.0863 & &0.0862& &  0.0866 && 0.0869  && 0.0862\\
		& &0.1054 & & 0.0608& & &0.0116 & &0.0115& &  0.0116 && 0.0117 && 0.0115\\
		\hline\\
		(40,15)& I&0.7871 & &0.1412& & &0.0805 & & 0.0804& &  0.0806 && 0.0807  && 0.0805\\
		& &0.5675 & &0.0313& & &0.0102 & &0.0101& &   0.0101 &&  0.0102  && 0.0102 \\
		& II&0.1807 & &0.1601& & &0.0833 & &0.0831& &  0.0834 && 0.0835  && 0.0832\\
		& &0.0809 & &0.0412& & &0.0109 & &0.0108& &  0.0109 && 0.0109  && 0.0108\\
		& III&0.1844 & &0.1709& & &0.0846 & & 0.0845& &   0.0847 && 0.0849  && 0.0846\\
		& &0.0591 & &0.0470& & &0.0113 & &0.0112& & 0.0112 && 0.0113 && 0.0113\\
		\hline
	\end{tabular}}
\end{table}

\begin{table}[!htbp]
	\renewcommand\thetable{3}
	\scriptsize \caption{\label{Table 3}Average length and coverage probability of $95\%$ C.I. for
	$\alpha$ and $\beta$ when $T=1.5.$}
	\begin{tabular}{cccccccccccccccccc}
		\hline
		& & & && $\alpha$ && & & & &&& &$\beta$\\
		\hline\\
		(n,m)& scheme &  ACI && boot-p&& boot-t&& HPD& & & ACI && boot-p&& boot-t&& HPD \\
		\hline\\
		(30,10)&I&  1.6866 && 2.4919&& 2.6417&& 0.8570& & & 0.8230 && 1.2655&& 1.6682&& 0.4053 \\
		& &   0.9283 && 0.8129&& 0.8295&& 0.9480& & & 0.8701 && 0.8653&& 0.9459&& 0.9492 \\
		& II&  1.5017 && 2.0124&& 2.2382&& 0.8601& & & 0.8981 && 1.0208&& 1.5391&& 0.4114 \\
		& &   0.9257 && 0.8677&& 0.8739&& 0.9487& & & 0.8871 && 0.9006&& 0.9095&& 0.9498 \\
		& III& 1.4682 && 1.9503&& 1.9832&& 0.8584& & & 0.9140 && 1.0408&& 1.3678&& 0.4171\\
		& & 0.9210 && 0.8779&& 0.8934&& 0.9473& & & 0.8892 && 0.9059&& 0.9190&& 0.9496 \\
		\hline\\
		(30,15)&I&  1.2722 && 1.6068&& 1.8069&& 0.8157& & & 0.7057 && 1.1223&& 1.3871&& 0.3992 \\
		& &  0.9426 && 0.8615&& 0.8725&& 0.9515& & & 0.9081 && 0.9201&& 0.9234&& 0.9401 \\
		& II &  1.2325 && 1.5764&& 1.7768&& 0.8344& & & 0.7489 && 0.8629&& 1.2013&& 0.4049 \\
		&  & 0.9194  && 0.8924&& 0.9227&&  0.9483& & & 0.9033 && 0.9199&& 0.9305&& 0.9492 \\
		& III&  1.1782 &&  1.4872&& 1.5273&& 0.8353& & & 0.7883 && 0.8960&& 1.0720&& 0.4113 \\
		&  &  0.9064 && 0.9018&& 0.9186&& 0.9476& & & 0.9055 && 0.9178&& 0.9621&& 0.9490 \\
		\hline\\
		(40,10)& I&  1.7059 && 2.4172&& 2.6725&& 0.8122& & & 0.8322 && 1.2006&& 1.9418&& 0.3977 \\
		& &  0.9360 && 0.7936&& 0.8493&& 0.9481& & & 0.8685 && 0.8387&& 0.9353&& 0.9496 \\
		& II& 1.4316 && 1.8620&& 1.9062&& 0.8151& & & 0.8971 && 0.9866&& 1.2241&& 0.4120 \\
		& &  0.9258 && 0.8656&& 0.9065&& 0.9475& & & 0.8842 && 0.8930&&  0.9019&& 0.9492 \\
		& III&  1.3989 && 1.8115&& 1.8417&&  0.8209& & & 0.9155 && 1.0130&& 1.2131&& 0.4154 \\
		& &  0.9238 && 0.8702&& 0.8926&& 0.9485& & & 0.8876 && 0.8987&& 0.9104&& 0.9492 \\
		\hline\\
		(40,15)& I& 1.2829 && 1.6239&& 1.8337&& 0.7890& & & 0.6861 && 0.8595&& 0.8488&&  0.3908 \\
		& & 0.9388 && 0.8591&& 0.8994&& 0.9501& & & 0.9026 && 0.9117&& 0.9477&& 0.9501 \\
		& II&  1.1784 && 1.4744&& 1.6841&&  0.7951& & & 0.7468 && 0.8402&& 1.0552&& 0.4038 \\
		& &  0.9185 && 0.8963&& 0.9163&& 0.9488& & & 0.9028 && 0.9159&& 0.9350&& 0.9498 \\
		& III&  1.1314 && 1.4043&& 1.6104&& 0.7951& & & 0.7833&& 0.8786&&  1.0143&& 0.4103 \\
		& & 0.9077 &&  0.9041&&  0.9183&& 0.9483& & & 0.9033 && 0.9138&& 0.9368&& 0.9598 \\
		\hline
\end{tabular}
\end{table}	
\begin{table}[!htbp]
	\renewcommand\thetable{4}
	\scriptsize \caption{\label{Table 4}Absolute bias and MSE of
		estimates of $\alpha$ when $T=0.75.$}
	\scalebox{0.95}{
	\begin{tabular}{cccccccccccccccccc}
		\hline\\
		(n,m)& scheme  &$\widehat{\alpha}$& &$\widehat{\alpha}_{MPS}$  && &&  $\widehat{\alpha}_{LI}$& & &  &$\widehat{\alpha}_{GE}$& &  &$\widehat{\alpha}_{SE}$ \\
		\hline	
		& & & && & &$p=-0.25$ & &$p=0.25$& &  $q=-0.25$ && $q=0.25$  && \\
		\hline\\
		(30,10)& I&0.4871 & &0.3250& & &0.1777 & &0.1764& &  0.1770 && 0.1772  && 0.1771\\
		& &1.2142 & &0.1950& & &0.0495 & &0.0483& &  0.0485 && 0.0484  && 0.0488\\
		& II&0.3655 & &0.2971& & &0.2001 & &0.2016& &  0.2039 && 0.2062  && 0.2008\\
		& &0.2699 & &0.1567& & &0.0607 & &0.0615& &  0.0615 && 0.0645  && 0.0611\\
		& III&0.3437 & &0.2863& & &0.1959& &0.1972& & 0.1994 && 0.2015  && 0.1965\\
		& &0.2349 & &0.1476& & &0.0581 & & 0.0588 &&  0.0601 &&  0.0613  && 0.0584\\
		\hline\\
		(30,15)& I&0.3245  & &0.2566& & &0.1707 & &0.1691& &  0.1696 && 0.1695  && 0.1699\\
		& &0.3725  & &0.1069& & &0.0456 & &0.0443& &  0.0445 && 0.0442 && 0.0449\\
		& II&0.3572 & &0.4163& & &0.2278 & &0.2314& & 0.2349 && 0.2386  && 0.2296\\
		& &0.2207 & &0.2267& & &0.0768 & &0.0791& & 0.0816 && 0.0842 && 0.0779\\
		& III&0.2671  & &0.3230& & &0.1907 & &0.1924& &  0.1945 && 0.1966  && 0.1915\\
		& &0.1248  & &0.1439& & &0.0543& &0.0550& &  0.0562 &&  0.0573  && 0.0546\\
		\hline\\
		(40,10)&I&0.4769 & &0.3398& & &0.1676 & &0.1661& &  0.1664 && 0.1663  && 0.1668\\
		& &0.9496 & &0.2050& & &0.0448 & &0.0435& &  0.0436 && 0.0433  && 0.0441\\
		& II&0.3451 & &0.2880& & &0.1874 & &0.1885& &  0.1904 && 0.1921  &&0.1879 \\
		& &0.2222 & &0.1429& & &0.0535 & &0.0539& & 0.0550 && 0.0560  && 0.0537\\
		& III&0.3308 & & 0.2794& & &0.1823 & &0.1832& &  0.1849 && 0.1865  && 0.1827\\
		& &0.2135 & &0.1412& & &0.0501 & &0.0504& &  0.0119 && 0.0118 && 0.0120\\
		\hline\\
		(40,15)&I&0.3420 & &0.2571& & &0.1639 & &0.1625& &  0.1629 && 0.1628  && 0.1632\\
		& &0.5346 & &0.1084& & &0.0428 & &0.0416& &  0.0417 && 0.0414  && 0.0422\\
		& II&0.3347 & &0.3945& & &0.2135 & & 0.2164& &  0.2192 && 0.2222 && 0.2149\\
		& &0.1797 & &0.2064& & &0.0672 & &0.0689& &  0.0708 && 0.0727  && 0.0681\\
		& III&0.2593 & &0.3116& & &0.1785 & &0.1798& &  0.1815 && 0.1833  && 0.1791\\
		& &0.1121 & &0.1331& & & 0.0479 & &0.0485& &  0.0494 && 0.0502 && 0.0482\\
		\hline
	\end{tabular}}
\end{table}
\begin{table}[!htbp]
	\renewcommand\thetable{5}
	\scriptsize \caption{\label{Table 5}Absolute bias and MSE of
		estimates of $\beta$ when $T=0.75$.}
	\scalebox{0.95}{
	\begin{tabular}{cccccccccccccccccc}
		\hline\\
		(n,m)& scheme  &$\widehat{\beta}$& &$\widehat{\beta}_{MPS}$  && &&  $\widehat{\beta}_{LI}$& & &  &$\widehat{\beta}_{GE}$& &  &$\widehat{\beta}_{SE}$ \\
		\hline	
		& & & && & &$p=-0.25$ & &$p=0.25$& &  $q=-0.25$ && $q=0.25$  && \\
		\hline\\
		(30,10)& I&0.7260 & &0.1730& & &0.0828 & &0.0827& &  0.0831 && 0.0834 && .0828\\
		& &0.5283 & & 0.0466& & & 0.0107 & & 0.0106& &  0.0107 &&  0.0108 &&  0.0107\\
		& II&0.2716 & &0.2289& & &0.0881 & &0.0877& &  0.0877 && 0.0875  && 0.0879\\
		& &0.1582 & &0.0803& & &0.0123 & &0.0122& & 0.0122 && 0.0121 && 0.0123\\
		& III&0.2475 & &0.2145& & &0.0865 & &0.0862& &  0.0862 && 0.0861  && 0.0864\\
		& &0.1259 & &0.0716& & &0.0117 & &0.0116& & 0.0116&& 0.0115  && 0.0117 \\
		\hline\\
		(30,15)& I&0.2638  & &0.1406& & &0.0819 & &0.0818& &  0.0822 && 0.0824 && 0.0819\\
		& &0.3930  & &0.0316& & &0.0105 & &0.0104& &  0.0105 && 0.0106  && 0.0105\\
		& II&0.2615 & &0.2453& & &0.0873 & &0.0867& &  0.0863 && 0.0859  && 0.0870\\
		& &0.1399 & &0.0910& & &0.0121 & &0.0119& &  0.0118&& 0.0116  && 0.0120\\
		& III&0.1886  & &0.1797& & &0.0831 & &0.0827& &  0.0827 && 0.0825  && 0.0829\\
		& &0.0649  & &0.0516& & &0.0109 & &0.0108& &  0.0108 && 0.0107 && 0.0109\\
		\hline\\
		(40,10)&I&0.3957 & &0.1840& & &0.0817 & &0.0816& &  0.0819 && 0.0821  && 0.0817\\
		& &0.1709 & &0.0520& & &0.0104 & &0.0103& &  0.0104 && 0.0105  && 0.0104\\
		& II&0.2683 & &0.2361& & &0.0885 & &0.0881& &  0.0880 && 0.0878  && 0.0883\\
		& &0.1397 & &0.0859& & &0.0122 & &0.0121& &  0.0121 && 0.0120  && 0.0122\\
		& III&0.2532 & &0.2250& & &0.0870 & &0.0867& &  0.0867 && 0.0866  && 0.0869\\
		& &0.1260 & &0.0783& & &0.0120 & &0.0119& &  0.0119 && 0.0118  && 0.0120\\
		\hline \\
		(40,15)&I&0.6908 & &0.1411& & &0.0798 & &0.0797& &  0.0800 && 0.0802  && 0.0797\\
		& &0.3047 & &0.0312& & &0.0100 & &0.0099& &  0.0100 && 0.0101  && 0.0100\\
		& II&0.2547 & &0.2507& & &0.0893 & &0.0887& &  0.0882 && 0.0877  && 0.0890\\
		& &0.1159 & &0.0963& & &0.0126 & &0.0123& &  0.0122 && 0.0120  && 0.0124\\
		& III&0.1914 & &0.1879& & &0.0829 & &0.0825& &  0.0825 && 0.0823 && 0.0827\\
		& &0.0669 & & 0.0557& & &0.0108 & &0.0107& &  0.0107 && 0.0106 && 0.0108\\
		\hline
	\end{tabular}}
\end{table}

\begin{table}[!htbp]
	\renewcommand\thetable{6}
	\scriptsize \caption{\label{Table 6}Average length and coverage probability of $95\%$ C.I. for
		$\alpha$ and $\beta$ when $T=0.75.$}
	{
	\begin{tabular}{cccccccccccccccccc}
		\hline
		& & & && $\alpha$ && & & & &&& &$\beta$\\
		\hline\\
		(n,m)& scheme &  ACI && boot-p&& boot-t&& HPD& & & ACI && boot-p&& boot-t&& HPD \\
		\hline\\
		(30,10)& I& 1.6775 && 2.0947&& 2.3369&& 0.8511& & & 0.8240 && 1.3510&& 1.5282&& 0.4020 \\
		& & 0.9348 && 0.8235&& 0.8749&& 0.9479& & & 0.8799 && 0.8763&& 0.9409&& 0.9491 \\
		& II& 1.3180 && 1.7148&& 1.9864&& 0.8880& & & 0.9686 && 1.2647&& 1.5943&& 0.4233 \\
		& &  0.8184 &&  0.8494&& 0.8962&& 0.9483& & & 0.8686 && 0.8495&& 0.9414&& 0.9490 \\
		& III&  1.2966 && 1.6360&& 1.8635&& 0.8649& & & 0.9847 && 1.1818&& 1.4015&& 0.4133 \\
		& &  0.8392 && 0.8899&& 0.9162&& 0.9492& & & 0.8970 && 0.8661&& 0.9239&& 0.9484 \\
		\hline\\
		(35,15)& I& 1.2752 && 1.6161&& 1.8547&& 0.8144& & & 0.7048 && 1.1695&& 1.3268&& 0.3988 \\
		& & 0.9392 && 0.8621&& 0.9034&& 0.9501& & & 0.9087 && 0.9198&& 0.9255&& 0.9535 \\
		& II&  1.0247 && 1.2023&& 1.3168&& 0.8723& & &  0.8260 && 1.0679&& 1.2178&&  0.4121 \\
		& &  0.9053 && 0.8429&& 0.8936&& 0.9493& & & 0.8475 && 0.8310&& 0.8752&& 0.9494 \\
		& III&  1.0743 && 1.1932&& 1.3591&& 0.8290& & & 0.8296 && 0.9052&& 1.0084&& 0.4019 \\
		& &  0.8648 && 0.8968&& 0.9106&& 0.9481& & &  0.9464 && 0.8820&&  0.9293&& 0.9495 \\
		\hline\\
		(40,10)&I&  1.4135 && 1.7339&& 1.8531&& 0.8191& & & 0.8299 && 1.2996&& 1.5237&& 0.3939 \\
		& &  0.9331 && 0.8825&& 0.9146&& 0.9485& & & 0.8541 && 0.8267&& 0.9317&& 0.9499 \\
		& II& 1.2681 && 1.6423&& 1.7529&& 0.8374& & & 0.9665 && 1.2478&& 1.3757&& 0.4231 \\
		& &  0.8226 && 0.8490&& 0.9205&& 0.9490& & & 0.8555 && 0.8464&& 0.9344&& 0.9488 \\
		& III&  1.2541 && 1.5694&& 1.7325&& 0.8137& & & 0.9796 && 1.1670&& 1.3149&& 0.4213 \\
		& &  0.8426 && 0.8839&& 0.9314&& 0.9478& & & 0.8831 && 0.8698&& 0.9208&& 0.9494 \\
		\hline\\
		(40,15)&I&  1.2852 && 1.6241&& 1.7638&& 0.7915& & & 0.6868 && 1.0593&& 1.2304&& 0.3885 \\
		& &  0.9434 && 0.8604&& 0.9126&& 0.9501& & & 0.9056 && 0.9154&& 0.9415&& 0.9571 \\
		& II&   0.9953&& 1.1725&& 1.2621&& 0.8439& & & 0.8193 && 1.0496&&  1.0636&& 0.4144 \\
		& &  0.8567 && 0.8665&& 0.9021&& 0.9495& & & 0.8337 && 0.9137&& 0.9311&& 0.9575 \\
		& III&  1.0347 &&  1.1417&& 1.2136&& 0.7788& & & 0.8264 && 0.9005&& 0.9658&& 0.4015 \\
		& & 0.8656 && 0.8887&& 0.9054&& 0.9475& & & 0.9426 && 0.8857&& 0.9653&& 0.9495 \\
		\hline
	\end{tabular}}
	\end{table}

\section{Real data}
In this section, a real data set is considered to illustrate the established estimates. Gumbel type-II distribution can be used as an alternative to some well known two parameter distributions such as Burr III , Nandrajah Haghighi (NH) and Inverted Kumaraswamy (Ikum) distributions. For the purpose of goodness of fit test, negative log-likelihood criterion, Alkaikes-information criterion (AIC), Bayesian information criterion (BIC), Cramer-von Mises $(C^{*})$ measure and Anderson-Darling $(A^{*})$ measure along with $p$-values are computed and tabulated in Table \ref{Table 7}. If we get value of $C^{*}$ and $A^{*}$ smaller but $p$-values greater, then the distribution fits better than other one. \\\\
 \textbf{Data : Death rates due to Covid-19 in India.} \\
 This following data set is taken from https://www.worldometers.info/coronavirus/country/india/, which represents the death rates due to Covid-19 in India from March $16$ to May $13$, $2020$. The data set is provided in the following: \\[0.2 cm]
 -----------------------------------------------------------------------------------------------------------------------\\
 13.33, 17.65, 17.65, 16.67, 17.86, 17.86, 22.58, 22.73, 20, 21.82, 30.77, 21.51, 22.22, 22.13, 23.88, 22.15, 28.16, 27.38, 30.94, 30.18, 26.46, 26.16, 25.48, 26.02, 26.33, 24.34, 22.91, 23.46, 23.26, 22.43, 21.87, 20.22, 19.23, 17.46, 16.38, 15.32, 13.96, 13.48, 12.58, 12.43, 12.20, 11.90, 11.63, 11.51, 11.34, 11.29, 10.89, 10.90, 10.57, 10.87, 10.69, 10.43, 10.12, 9.99, 9.82, 9.54, 9.23, 9, 8.81, 8.65, 8.34, 7.74, 7.60, 7.45, 7.24, 7.03, 6.87, 6.71, 6.64, 6.52, 6.43, 6.33, 6.27,6.23,5.68,5.63,5.56, 5.53, 5.49, 5.53, 5.54, 5.55, 5.53, 5.50, 5.47, 5.44, 5.44, 5.47, 5.45, 5.36.\\
 -----------------------------------------------------------------------------------------------------------------------\\[0.2 cm]
 For the purpose of goodness of fit test, we consider different plots in Figure $3$, Figure $4$ and Figure $5$. In Figure $3$, the theoretical CDFs of the distributions are compared with the empirical CDF of the given real data set. In Figure $4$, the QQ-plots are used to compare the fitted theoretical models. In Figure $5(a)$, the box plot of the real data set is displayed, which represents that the distribution is right skewed. The TTT plot for given real data set is shown by Figure $5(b)$.  From Table \ref{Table 7}, it is concluded that Gumbel type-II distribution is fitted to that data better than NH, Burr III and IKum distributions.
 From the real data set, different AT-II progressive hybrid censored samples are considered by using different values of $n,~m$ and $T$. The computed values of the MLEs, MPSEs and Bayes estimates based on the real data set are tabulated in Table \ref{Table 8}. Further, the confidence intervals are constructed and tabulated in Table \ref{Table 9}. From Table \ref{Table 8}, it is observed that the Bayes estimates perform better than the other estimates. From Table \ref{Table 9}, it is seen that the HPD credible interval performs better than asymptotic confidence intervals.

\begin{figure}[h!]
	\begin{center}
		\subfigure[]{\label{c1}\includegraphics[height=2in]{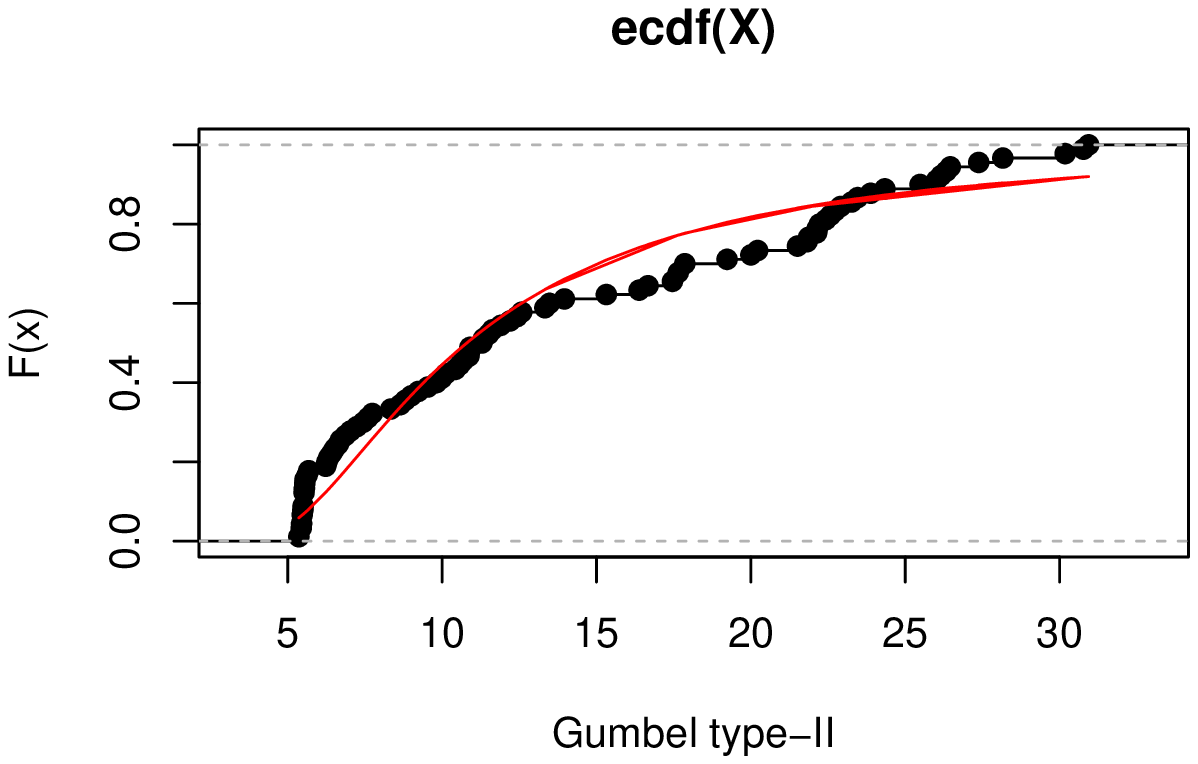}}
		\subfigure[]{\label{c1}\includegraphics[height=2in]{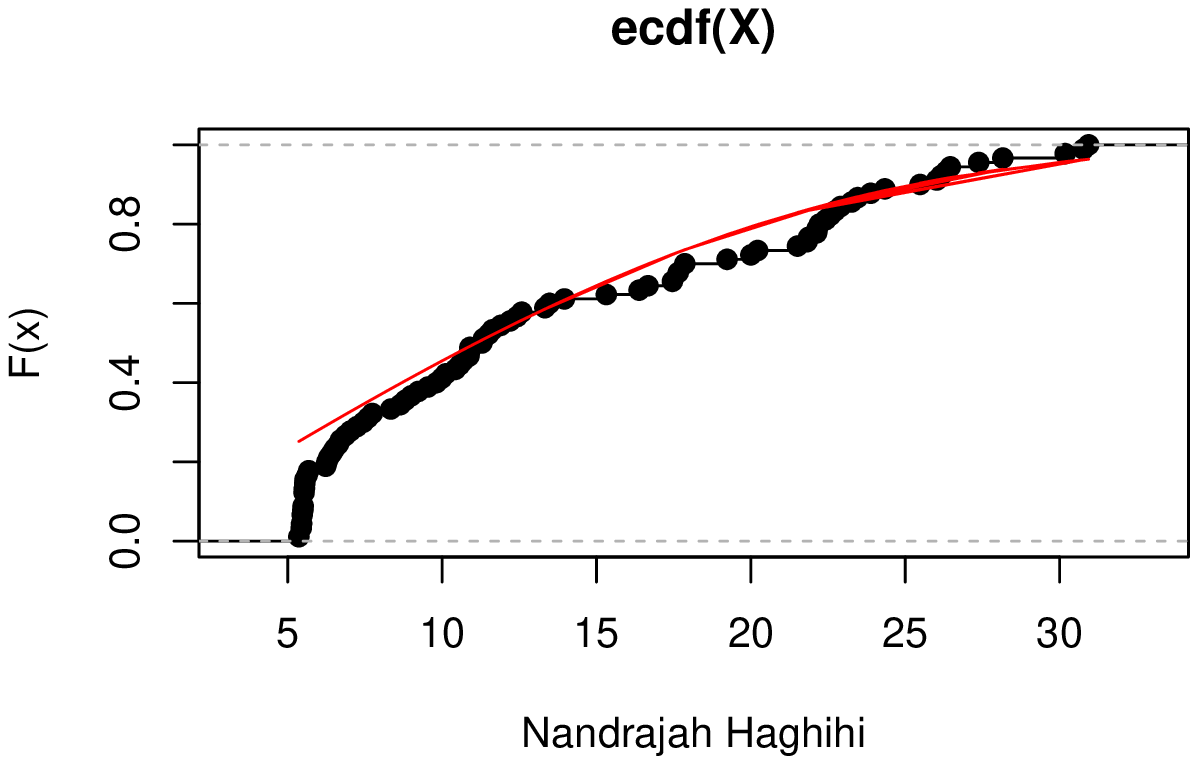}}
		\subfigure[]{\label{c1}\includegraphics[height=2in]{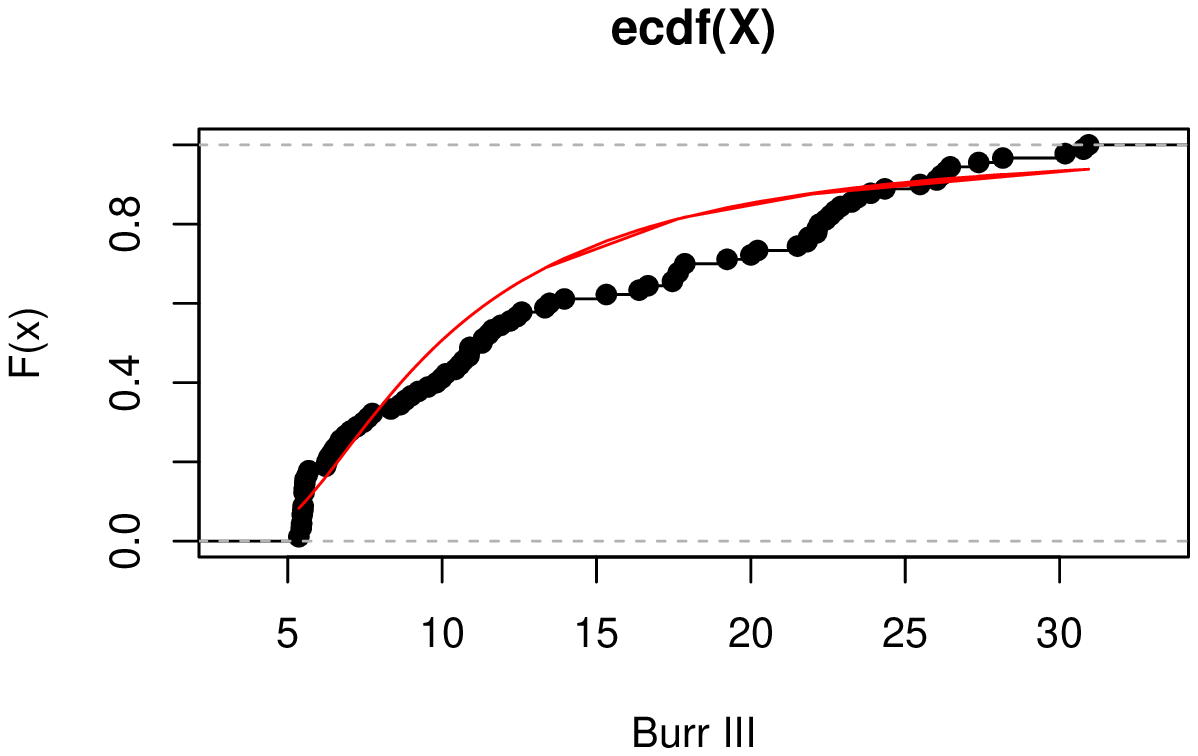}}
		\subfigure[]{\label{c1}\includegraphics[height=2in]{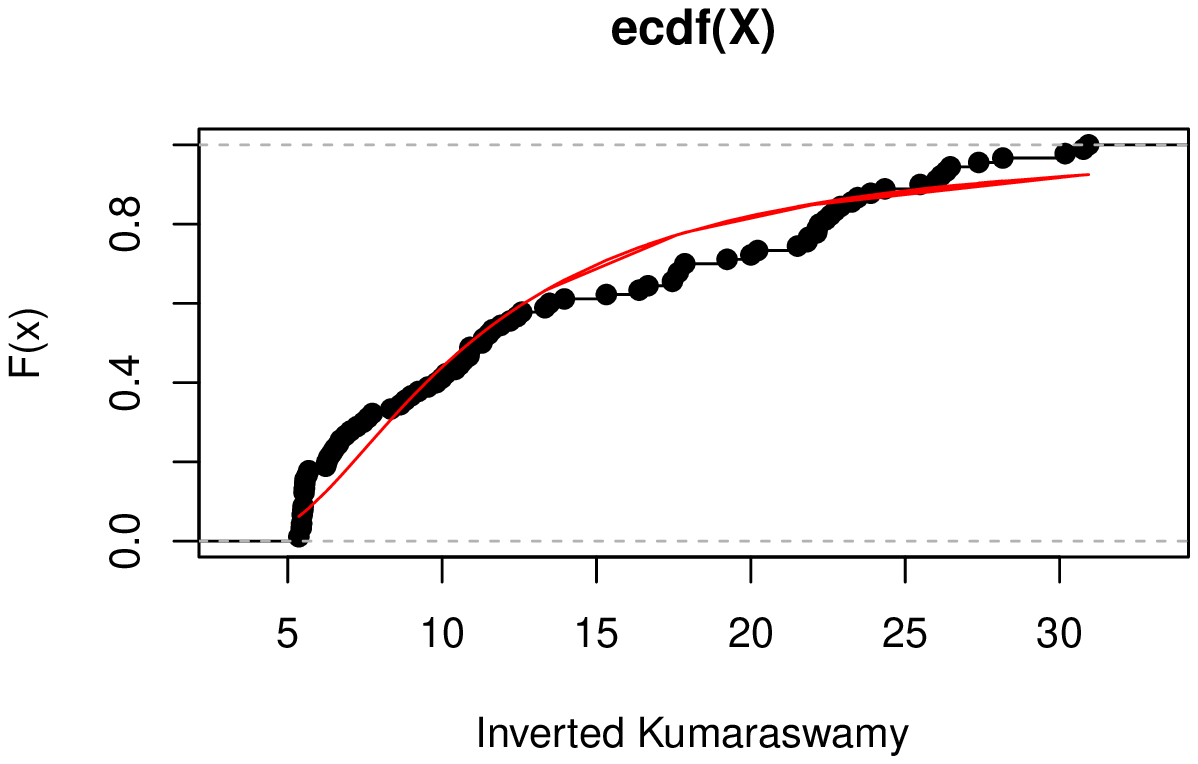}}
\caption{ECDF and CDF comparison for different distributions fitted to given real data set. }
\end{center}
\end{figure}

\begin{figure}[h!]
	\begin{center}
		\subfigure[]{\label{c1}\includegraphics[height=2in]{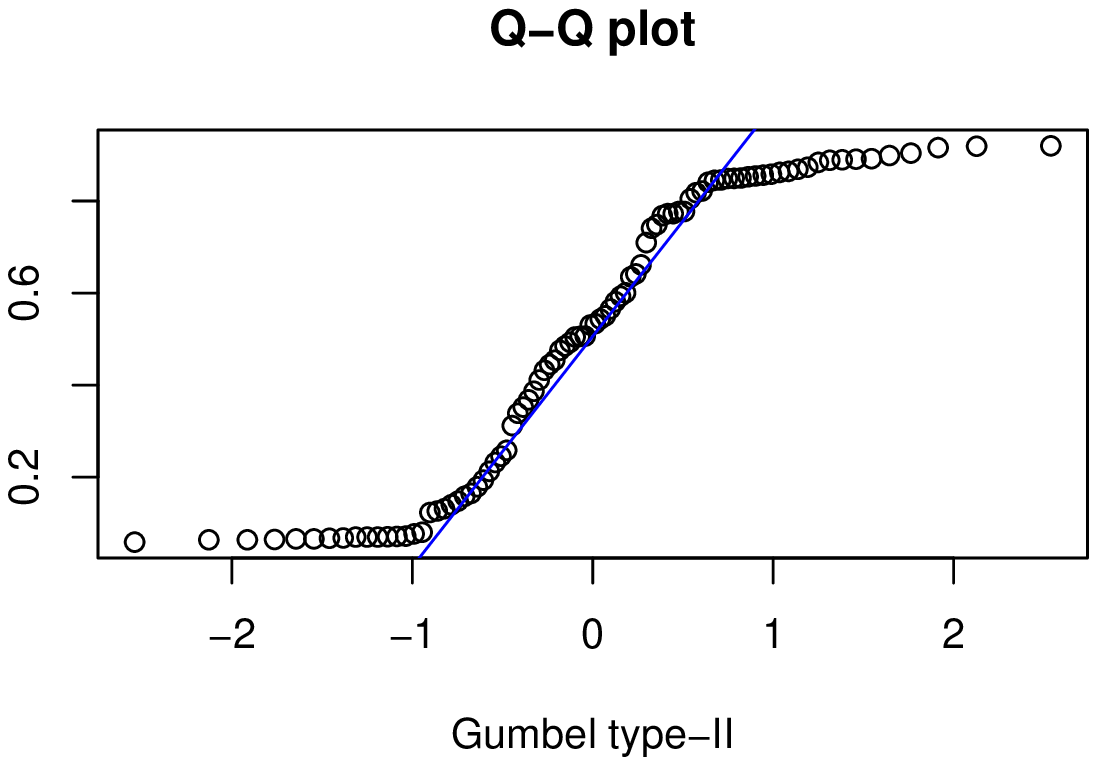}}
		\subfigure[]{\label{c1}\includegraphics[height=2in]{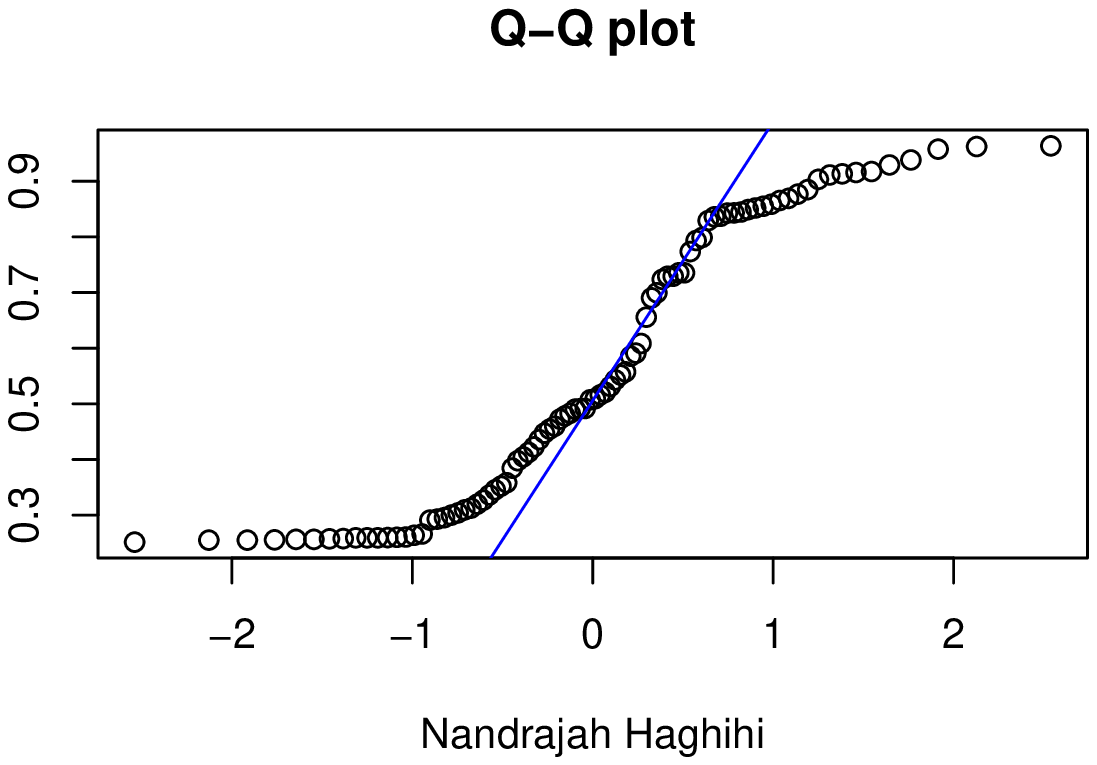}}
		\subfigure[]{\label{c1}\includegraphics[height=2in]{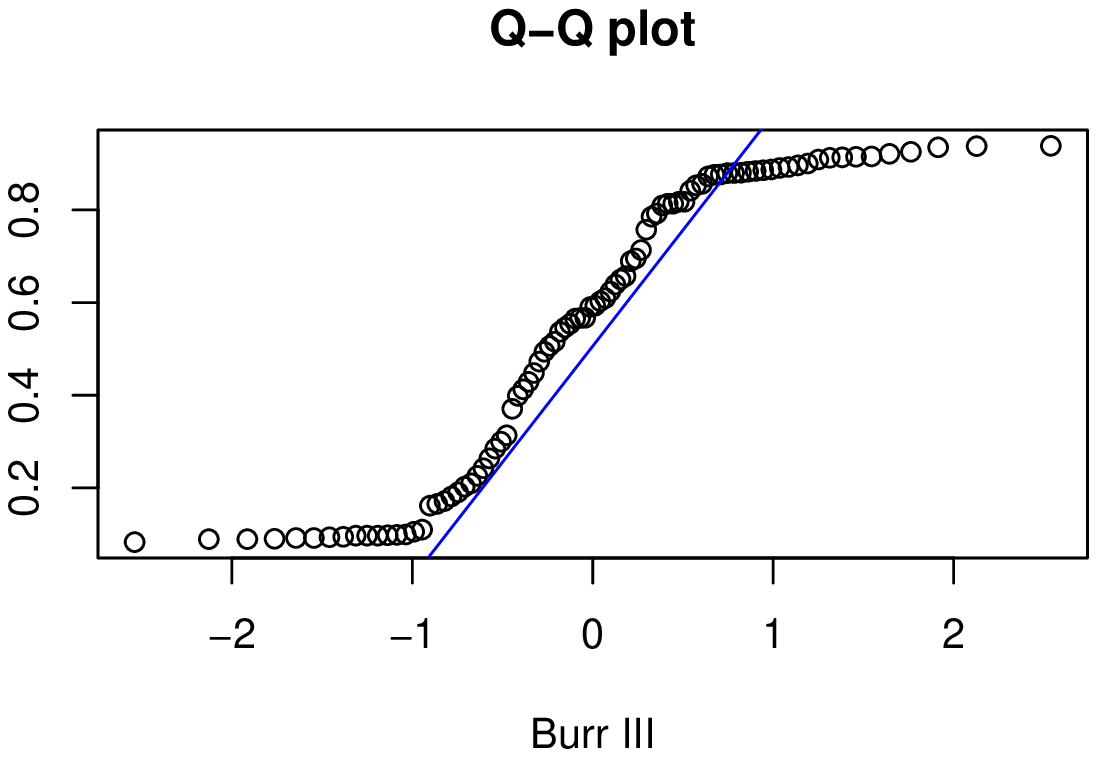}}
		\subfigure[]{\label{c1}\includegraphics[height=2in]{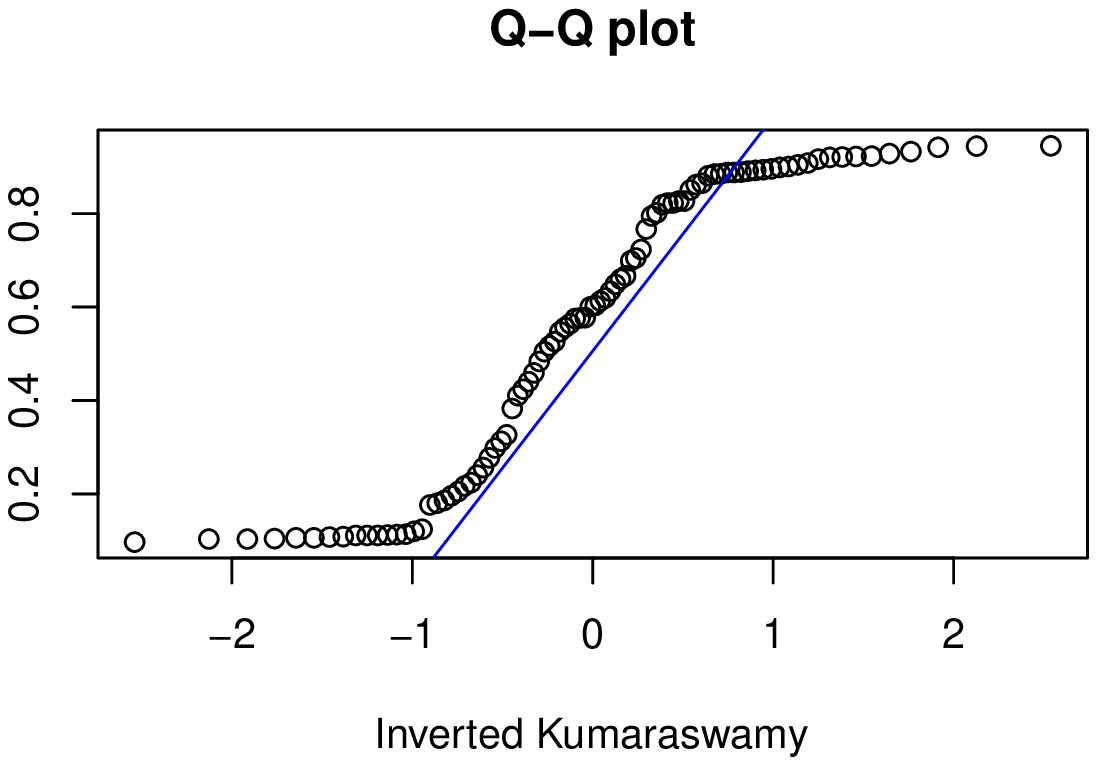}}
		\caption{QQ-plot comparison for different distributions fitted to given real data set.  }
	\end{center}
\end{figure}

\begin{figure}[h!]
	\begin{center}
      \subfigure[]{\label{c1}\includegraphics[height=1.5in,width=3in]{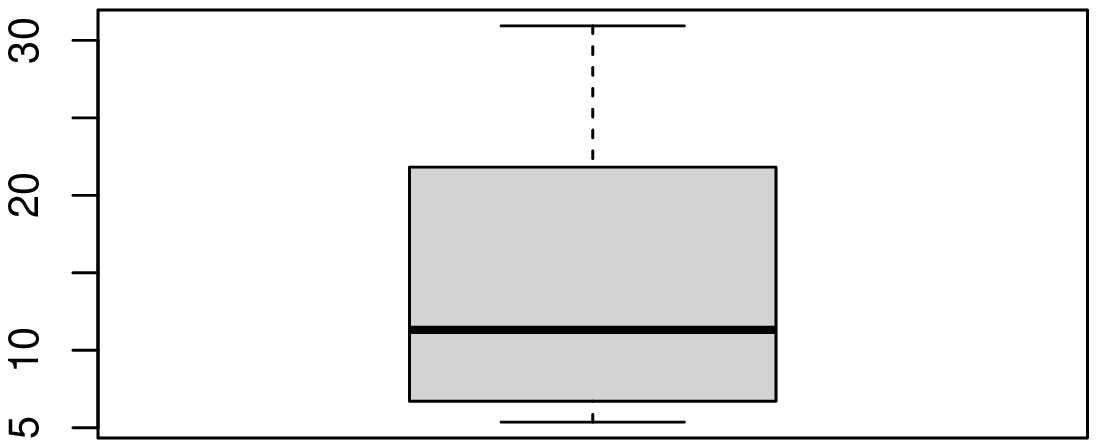}}
      	\subfigure[]{\label{c1}\includegraphics[height=1.5in,width=3in]{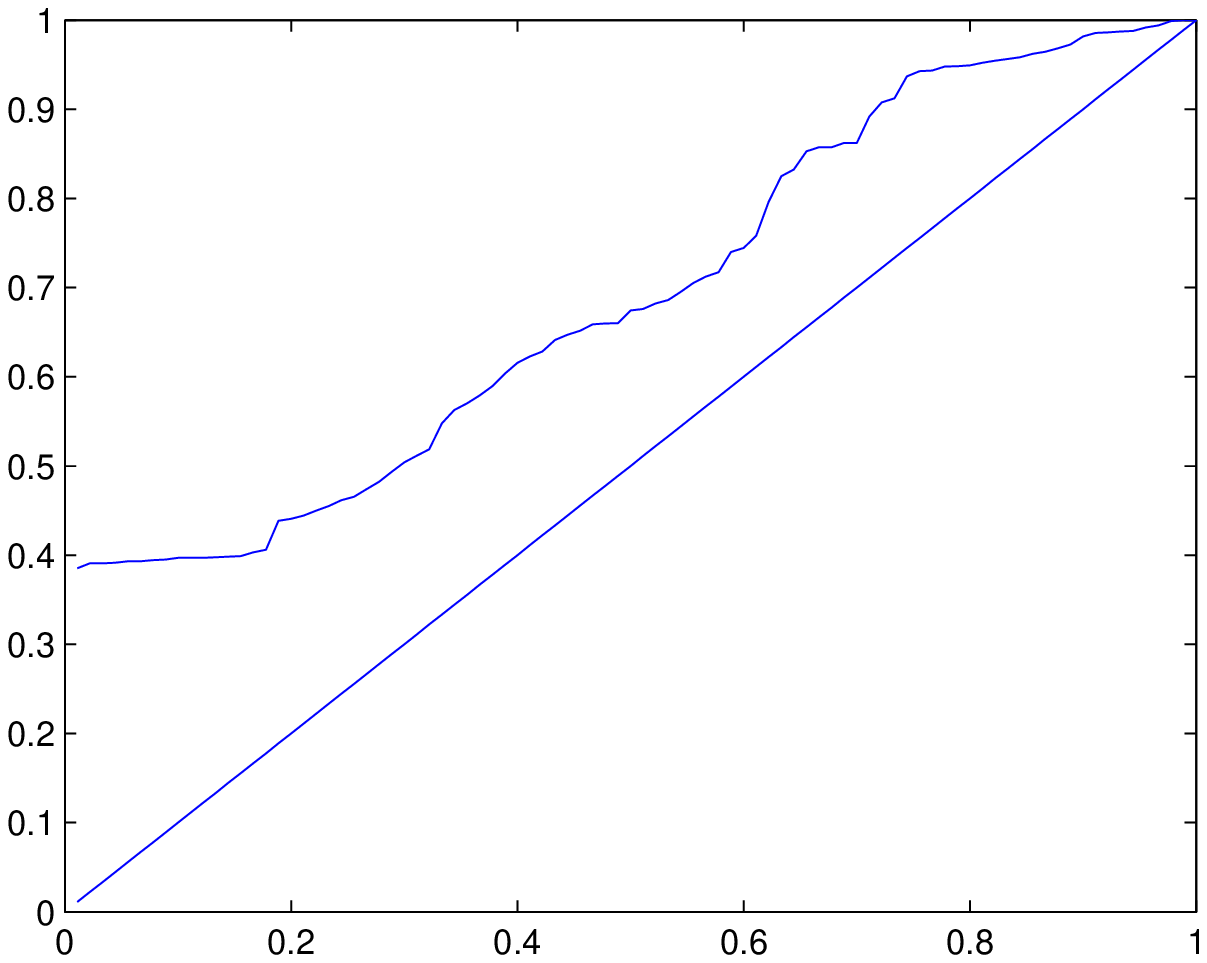}}
      \caption{(a) Boxplot and (b) TTT plot for given real data set.}
	\end{center}
\end{figure}

\begin{table}[!htbp]
	\renewcommand\thetable{7}
	\scriptsize \caption{\label{Table 7}The values of MLEs and statistics of different distributions along with goodness-of-fit measures for Covid-19 data set.}
	\begin{center}
	\begin{tabular}{cccccccccccccccccc}
		\hline \\
	Model&& $\widehat{\alpha}$&&$\widehat{\beta}$& & -$\log L$ & AIC & BIC& $C^{*}$ &  $A^{*}$& p-value \\
	\hline\\
	GT-II &&	2.0130&& 82.7737&& 300.6597& 605.3194 & 610.3190& 0.1694 &1.2297 &0.9674\\
     NH &&   138.7024 && 0.0003 && 311.8292& 627.6584& 632.6580& 0.2528 &1.4784 &0.8398\\
     Burr III && 2.0256&& 85.8196&& 300.7166& 605.4332& 610.4328&0.3204 &1.8480 & 0.6588\\
     IKum && 2.2073&& 163.2839&& 300.6774& 605.3548& 610.3544&0.2219 &1.3392 & 0.8927\\
     \hline\\
	\end{tabular}
\end{center}
\end{table}	

\begin{table}[!htbp]
	\renewcommand\thetable{8}
	\scriptsize \caption{\label{Table 8}Average values of the
		estimates of parameters for real data set.}
	\scalebox{0.9}{
	\begin{tabular}{cccccccccccccccccc}
		\hline\\
		(n,m)&T& scheme& $\theta$  &$MLE$& $MPSE$  &&  &$LLF$& &  &$GELF$&   &SELF \\
		\hline	
		& & & & & &   &$p=-0.25$ & &$p=0.25$&   $q=-0.25$ & &$q=0.25$  & \\
		\hline\\
		(90,40)& 10&(0*39,50)& $\alpha$&1.8701 &1.6657 &   &2.0298 & &2.0291&  2.0290 & &2.0286  & 2.0294\\
		& & & $\beta$&63.4539  &52.9911 &   &85.9635 & &85.9626& 85.9631 & &85.9630  & 85.9631\\
		& 15& & $\alpha$&1.9209 &  1.7090 &&2.0285 & &2.0280&   2.0279 && 2.0277 & 2.0282\\
		& & & $\beta$&68.4234  &56.2260  & &86.1488 & &86.1461&   86.1474 && 86.1475  & 86.1475\\
		& 10& (0*35,10*5)&$\alpha$& 1.9380 &1.7362 &   &2.0310 & &2.0298&   2.0295 & &2.0289  & 2.0304\\
		& & & $\beta$&71.7506 &60.2997 &   &86.0327 & &86.0322&  86.0325 & &86.0326  & 86.0325\\
		& 15& &$\alpha$&2.0281 &1.8208  & &2.0266 & &2.0257&  2.0255 && 2.0251  & 2.0262\\
		& & &  $\beta$&83.0597 &69.0278  & &85.9584 & &85.9576& 85.9580 && 85.9580  & 85.9581\\
		&10& (0*30,5*10)&$\alpha$&2.1209 &1.9269 &   &2.0449 & &2.0443& 2.0442 & &2.0439 &  2.0446\\
		& & & $\beta$&99.6236 &85.2831 &   &86.0510 & &86.0502& 86.0507 & &86.0506  & 86.0507\\
		&15 &&  $\alpha$&2.2134&2.0142  & &2.0692 & &2.0685 & 2.0683 && 2.0680 & 2.0688\\
		& & &$\beta$&115.7839 &98.1035  & &86.1114& &86.1075&   86.1094&& 86.1093  & 86.1095 \\
		\hline\\
		(90,50)&10& (0*49,40)& $\alpha$&1.7442 &1.5885 &   &1.9779 & &1.9773&    1.9772 & &  1.9769  &  1.9776\\
		& & & $\beta$&53.1119 &48.2612 &   &86.0341 & &86.0333&   86.0337 & & 86.0337  &  86.0337\\
		&15 && $\alpha$&1.9620 &1.7843  & &2.0283 & &2.0277&   2.0276 && 2.0273 & 2.0281\\
		& & &$\beta$ &73.7502 &63.7147  & &85.9503 & &85.9502& 85.9503 && 85.9501  & 85.9503\\
		& 10&(0*45,5*8)&$\alpha$&1.7442  &1.5885 &   &1.9682 & &1.9678&   1.9677 & &1.9674  & 1.9680\\
		& & & $\beta$&53.1119 & 48.2612 &   & 85.9022 & &85.8993& 85.9008 & &85.9007  & 85.9008\\
		& 15&& $\alpha$&2.0428 &1.8740  & &2.0410 & &2.0403&   2.0402 && 2.0399  & 2.0407\\
		& & &$\beta$&85.5752 &75.4799  & &85.8843 & &85.8827&  85.8835 &&  85.8834  &  85.8835\\
		&10&(0*40,4*10)& $\alpha$&1.7442 &1.5885 &   &1.9857 & &1.9851&  1.9850 & &1.9847  & 1.9854\\
		& && $\beta$&53.1119 &48.2612 &   & 85.8008 & &85.7989& 85.7998 & &85.7998  &85.7999 \\
		& 15&& $\alpha$&2.1158 &1.9539  & &2.0516 & &2.0510&   2.0508 && 2.0506  & 2.0513\\
		& & &$\beta$ &97.8510 &87.7249  & &86.2835 & &86.2772& 86.2803 && 86.2802 & 86.2804\\
		\hline\\
	\end{tabular}}
\end{table}
\begin{table}[!htbp]
	\renewcommand\thetable{9}
	\scriptsize \caption{\label{Table 9}Confidence intervals of parameters for real data set when $T=15$.}
	\begin{center}
	\begin{tabular}{cccccccccccccccccc}
		\hline\\
		& & & & & $\alpha$& & & & $\beta$ && \\
		\hline\\
		(n,m)& T& scheme&  &  ACI && HPD &&  ACI && HPD \\
		\hline\\
		(90,40)& 10& (0*39,50)& &(1.40,2.33) && (1.93,2.13) &&  (4.71,122.18) && (85.83,86.06) \\
		 &15& &  & (1.44,2.40) && (1.96,2.07) &&  (3.62,133.22) && (85.75,86.42) \\
		 &10& (0*35,10*5)& & (1.45,2.42) && (1.91,2.18) && (3.37,140.12) && (85.96,86.12) \\
		& 15&& & (1.52,2.53) && (1.90,2.12) &&  (0.78,165.33) && (85.87,86.06) \\
		& 10& (0*30,5*10)& &  (1.59,2.64) && (1.93,2.12) &&  (0,201.80) && (85.94,86.16) \\
		&  15&& &  (1.66,2.76) && (1.96,2.16) && (0,238.78) && (85.94,86.36) \\
       \hline\\
       (90,50)& 10&(0*49,40)& & (1.36,2.12) && (1.88,2.07) && (11.36,94.85) && (85.94,86.16) \\
       & 15&& & (1.53,2.39) && (1.93,2.10) &&  (9.94,137.55) && (85.89,86.01) \\
       & 10& (0*45,8*5)& & (1.36,2.12) && (1.88,2.04) &&  (11.36,94.85) && (85.76,86.11) \\
        & 15&& & (1.59,2.48) && (1.92,2.13) && (9.22,161.92) && (85.74,86.01) \\
        & 10&(0*40,4*10)& & (1.36,2.12) && (1.89,2.07) && (11.36,94.85) && (85.68,86.15) \\
        & 15&& & (1.65,2.57) && (1.95,2.13) &&  ( 7.96,187.74) && (85.91,86.49) \\
        \hline\\
\end{tabular}
\end{center}
\end{table}	
\section{Conclusion}
In this article, different estimates of the unknown model parameters of the Gumbel type-II distribution based on AT-II PHCS have been developed. Both classical and Bayesian estimation methods are used to obtain the estimates. It is observed that the MLEs and MPSEs can not be obtained explicitly. So, Newton Raphson iterative method is employed to compute these estimates. Bayes estimates are obtained based on the symmetric and asymmetric loss functions under the assumption of independent gamma priors using MCMC method. Three types of confidence intervals for unknown parameters are  constructed. Then, Monte carlo simulation study is performed to compare the performance of the estimates in terms of the absolute bias and MSEs. It is observed that the Bayes estimates under LINEX loss function perform better than the other estimates. Further, a real data set is considered for illustrative purposes.\\ [0.3 cm]
\textbf{Acknowledgements:}
The author S. Dutta, thanks the Council of Scientific and Industrial Research (C.S.I.R.
Grant No. 09/983(0038)/2019-EMR-I), India, for the financial assistantship received to carry out this
research work. Both the authors thanks the research facilities received from the Department of Mathematics, National Institute of Technology Rourkela, India.\\
\bibliography{myref1}
\end{document}